\documentclass[aip,pop,reprint,groupaddress,noshowpacs]{revtex4-1}
\pdfoutput=1

\usepackage{mathrsfs}
\usepackage{epsfig}
\usepackage{amsmath}
\usepackage{subfigure}
\usepackage{amsfonts}

\begin{document}

%
%

\title{Turbidity current flow over an obstacle and phases \\ of sediment wave generation}
%

%
%

\author{Moshe Strauss}
\affiliation{Department of Physics, Nuclear Research Center, Negev, Beer Sheva, Israel}
\affiliation{CSIRO, Kensington, WA, Australia}

\author{Michael E. Glinsky}
\affiliation{CSIRO, Kensington, WA, Australia}
\affiliation{School of Physics, University of Western Australia, Crawley, WA, Australia}

%
%

\begin{abstract}
We study the flow of particle-laden turbidity currents down a slope and over an obstacle.  A high-resolution 2D computer simulation model is used, based on the Navier-Stokes equations.  It includes poly-disperse particle grain sizes in the current and substrate.  Particular attention is paid to the erosion and deposition of the substrate particles, including application of an active layer model.  Multiple flows are modeled from a lock release that can show the development of sediment waves (SW).  These are stream-wise waves that are triggered by the increasing slope on the downstream side of the obstacle.  The initial obstacle is completely erased by the resuspension after a few flows leading to self consistent and self generated SW that are weakly dependant on the initial obstacle.  The growth of these waves is directly related to the turbidity current being self sustaining, that is, the net erosion is more than the net deposition.  Four system parameters are found to influence the SW growth:  (1) slope, (2) current lock height, (3) grain lock concentration, and (4) particle diameters.  Three phases are discovered for the system:  (1) ``no SW'', (2) ``SW buildup'', and (3) ``SW growth''.  The second phase consists of a soliton-like SW structure with a preserved shape.  The phase diagram of the system is defined by isolating regions divided by critical slope angles as functions of current lock height, grain lock concentration, and particle diameters.  
\end{abstract}


%
%

\maketitle

%
%

\section{Introduction}
Turbidity currents can trigger a variety of topographical behaviors by erosion and deposition over the sea floor, such as sediment waves (SW).  These currents are particle laden and gravity driven, where the particles are suspended by fluid turbulence  \citep{meiburg.kneller.10}.  When the bottom slope is large enough, the current can propagate in a self-sustained mode with increasing mass and high velocity  \citep{parker.etal.86, blanchette.etal.05, sequeiros.etal.09, pantin.franklin.09}.

Migrating SW generated by erosive turbidity currents have been reported in a variety of marine settings which include splays from submarine levees and submarine fans \citep{wynn.etal.00, wynn.stow.02}.  The time scale of SW formation can be thousands of years and include a sequence of many turbidity currents.  Typical SW wavelengths are in the range of 100 m to 5 km, and heights are in the range of 5 m to 100 m.  A series of turbidity currents flowing across a rough sea can form a field of SW that migrates upstream \citep{kubo.nakajima.02, lee.etal.02}.  A relation, $\lambda = 2 \pi h$ was found between the SW wavelength, $\lambda$, and the turbidity current height, $h$, that is in agreement with observation for typical $h = 60$~m and $\lambda = 380$~m \citep{normark.etal.80, wynn.stow.02}.

The traditional explanation of the mechanism for generating a train of up-streaming SW is based on sequence of turbidity currents flowing over an erodible bed.  A supercritical flow, where the kinetic energy of the flow is larger than the potential energy (Froude number larger than one), is considered a favorable condition for the formation.  An obstacle on the slope induces an erosion on the downstream side of the obstacle leading to a subsequent decrease in slope and to the formation of the next obstacle.  This establishes to a train of downstream crests in the waveform.  The upstream migration of the waveform results from the preferential deposition of sediment on the upslope and the preferential erosion on the downslope.  The generation of downstream undulations and the upstream migration by deposition and erosion can generate an extensive SW field.  This mechanism of generating SW is similar to the generation of transportational cyclic steps \citep{parker.izumi.00, taki.parker.05, sun.parker.05}.  Each cyclic step is bounded by a hydraulic jump and the resulting deposition and erosion causes the waves to migrate upstream.  

Numerical simulations have been carried out in order to explain the formation of SW by turbidity currents.  The models can be divided into two categories:  depth-averaged models, and depth-dependant models.  The Navier-Stokes depth-averaged models perform 1D simulations of turbidity currents flowing downslope over an erodible bed.  Pre-existing topography, such as surface roughness or a break in slope, are required to trigger the formation and growth of SW \citep{kubo.nakajima.02, fildani.etal.06, kostic.parker.06}.  These models do not include the eddy structure in the turbulent flow imprinted by the SW periodicity.  They also are unable to capture the detailed interaction between the sediment bed and the current close to the bed.

A linear stability analysis to generate SW based on the 2D depth dependent Navier-Stokes equations was done by \citet{hall.etal.08} and \citet{hall.09}.  Their results are consistent with a growth of the SW and their upstream migration.  There are approximations made in this analysis.  The front of the current is assumed to have passed so that the SW is growing underneath the body of the current.  An unrealistic erosion model without a threshold behavior is used.  Finally the expression used for the flow is not self consistent with the substrate structure.

We are motivated to eliminate the approximations used in these studies, and to obtain a more complete understanding of what controls the character of the SW generation.  We therefore study SW using a geometry and computer simulation method that takes into account the non-linearity, use a realistic erosion model, and model the depth dependant behaviour in a self consistent and self generating way.  In our treatment we apply non-linear simulations based on the 2D depth-dependent Navier-Stokes equations \citep{hartel.etal.00, necker.etal.02, blanchette.etal.05} with a realistic erosion relation \citep{garcia.parker.93, wright.parker.04, kostic.parker.06}.  The model includes the effects of poly-disperse particles in the current and in the substrate, and a sequence of flows with a self consistent coupling to the substrate.  An obstacle on the slope is used to trigger the possible buildup and possible growth of the SW.  The flows are initiated from a lock release.

We study the character of the SW generation as a function of four controlling parameters:  (1) slope, (2) current lock height, (3) grain lock concentration, and (4) particle diameters.  Three distinct phases of the SW generation are observed.  Regions of the controlling parameter space are identified for each of the phases -- the phase diagram.  The boundaries between these regions are directly related to the self-sustainment of the flow.  The first will be shown to be related to the flow being depositional everywhere on the slope, the second related to the flow being self sustaining only over the downslope part of the obstacle, and the third related to the flow being self sustaining everywhere.  The first condition results in no SW formation, the second in the formation of a soliton like structure and the third with growing SW.  The soliton like structure \citep{ablowitz.segur.84}, is a relatively constant periodic profile that migrates updip.  This stable profile exists at the threshold between deposition and self-sustainment.  The similarity to the buildup mode in a laser \citep{arecchi.etal.89}, led us to calling it the SW buildup phase.

The relationship of these solutions to the depth-averaged models is studied by forming depth-averaged variables from the detailed depth profiles.  A periodic structure in the flow is noted.  It is synchronised to the sediment waves in the substrate.  No such structure is seen in the depth-averaged models.  They have a very smooth character.  This is not surprising since they do not incorporate the eddy structure.

In the following sections we will present the physical model and numerical approach (Sec. \ref{model.desc}), followed by the simulations results (Sec. \ref{simulation.results}) and concluding remarks (Sec. \ref{conclusions}).

\section{Model description}
\label{model.desc}

\subsection{Governing equations}
We consider a particle-laden turbidity current model for which the particle concentration is relatively low ($\sim 1\%$) and the interaction between the particles can be ignored.  Hence, the density variation appears only in the gravity term (the Boussinesq approximation).  We assume that the particles are transported by the current and settle relative to the fluid in the direction of the gravity vector.  The system is assumed to be two-dimensional with normalized variables: $x=\tilde{x}/L_0$, $y=\tilde{y}/L_0$, and $t=\tilde{t}/t_0$, where $\tilde{x}$ and $\tilde{y}$ are the un-normalized space variables and $\tilde{t}$ is the un-normalized time variable.  Here, a characteristic length scale, $L_0$, is used and the time is normalized as $t_0=L_0/u_b$.  The buoyancy velocity is defined as
\begin{equation}
u_b=\sqrt{R_* c_0 g L_0},
\end{equation}
where $g$ is the gravity constant, $c_0$ is the initial grain concentration in the lock, $R_* = (\rho_p-\rho_f)/\rho_f$ is the fractional density difference, $\rho_p$ is the grain particle density, and $\rho_f$ is the fluid density.  The concentration of grain type $i$, $\tilde{c}_i$, is normalized to give $c_i = \tilde{c}_i / c_0$.  The variable $x$ is in the local flow direction, $y$ is in the perpendicular direction and $\theta$ is the local angle between the direction of gravity and the negative $y$ direction.  In order to model complex topgraphies we use a spatially varying gravity vector with and angle $\theta$ \citep{blanchette.etal.05, blanchette.etal.06}.  A curvilinear coordinate system is simulated with the second order curvatures being neglected.  This approximation is valid for flow heights smaller than the radius of curvature of the bottom topography.

The current equation, in normalized units, are written in terms of the vorticity, $\omega$, and the stream function, $\psi$,
\begin{equation}
\label{ux.eqn}
u_x = \frac{\partial \psi}{\partial y},
\end{equation}
\begin{equation}
u_y = - \frac{\partial \psi}{\partial x},
\end{equation}
\begin{equation}
\label{vorticity.eqn}
\omega = \frac{\partial u_y}{\partial x} -  \frac{\partial u_x}{\partial y},
\end{equation}
where $u_x = \tilde{u}_x/u_b$ and  $u_y = \tilde{u}_y/u_b$ are the normalized velocities.  These equations are consistent with the continuity equation
\begin{equation}
 \frac{\partial u_x}{\partial x} +  \frac{\partial u_y}{\partial y} = 0.
 \end{equation}
 The resulting current equations for $\omega$, $\psi$, and $c_i$ are \citep{hartel.etal.00, necker.etal.02, blanchette.etal.05} 
 \begin{equation}
 \label{lie.derivative.vorticity}
 \frac{\partial \omega}{\partial t} + (\vec{u} \cdot \nabla) \omega = \frac{1}{R_e} \nabla^2 \omega + (\hat{g} \times \nabla c )_z , 
\end{equation}
 \begin{equation}
 \label{stream.helmholtz}
\nabla^2 \psi = - \omega ,
\end{equation}
 \begin{equation}
 \label{lie.derivative.c}
\frac{\partial c_i}{\partial t} + (\vec{u}+u_{si} \hat{g})\cdot \nabla c_i = \frac{1}{P_e} \nabla^2 c_i ,
 \end{equation}
 where $c=\sum_{i} {c_i}$ is the normalized concentration of grains in the current (initially $c=1$), and $\hat{g} \equiv ( \sin{\theta},-\cos{\theta})$ is a unit vector in the direction of gravity.  Equation (\ref{lie.derivative.vorticity}) is obtained from the Navier-Stokes momentum equation and includes the turbulent motion of the flow.  Equation (\ref{stream.helmholtz}) is obtained from Eqs. (\ref{ux.eqn})--(\ref{vorticity.eqn}).  In these equations we have used the system's Reynolds number, $R_e = u_b L_0 / \nu$, where $\nu$ is the fluid viscosity.  The Peclet number $P_e = S_c R_e$ is related to the Schmidt number, $S_c = \nu/\kappa$, where $\kappa$ is the particle diffusion constant.  We assume that small scale unresolved flow structure will affect the transport of particles in the same way as the transport of the fluid, so we set $P_e = R_e$ or $S_c = 1$ \citep{hartel.etal.00}.  The settling velocity, $\tilde{u}_{si}$, for grain type $i$ is normalized to be $u_{si} = \tilde{u}_{si} / u_b$.  Note that the driving force of the current, in Eq. (\ref{lie.derivative.vorticity}), comes from the variation in the concentration $c$ perpendicular to $\hat{g}$.
 
The exchange of particles between the substrate and the current is governed by an Exner type equation for the substrate elevation $\eta(x,t)$ in accordance with the sediment transport rate \citep{parker.etal.00, pratson.etal.01},
\begin{equation}
\label{exner.eqn}
(1-\lambda_p) \frac{\partial \eta}{\partial t} = \sum_{i}{(J_{si}-J_{ri})} ,
\end{equation}
where $J_{si}$ and $J_{ri}$ are the volume rate of deposition and resuspension from the substrate surface for grain type $i$, respectively.  The sum in Eq. (\ref{exner.eqn}) is over all types of grains, $i$, and $\lambda_p$ is the average substrate porosity.  The substrate is divided into an upper and lower layer, where the upper layer is an active layer (AL) with thickness $L_a$.  Exchange of particles between the substrate and the current happens via this layer.

We use, for the current, a rectangular computational domain.  At the boundary, we enforce a non-slip, no normal flow condition, $\psi ={ \partial \psi}/{\partial y} = 0$, at the top and bottom boundaries.  We also impose a no normal flow condition at the left and right walls so that $\psi = {\partial^2 \psi}/{\partial x^2} = 0$.  This allows the use of fast Fourier transforms in the $x$ direction for high accuracy \citep{hartel.etal.00, blanchette.etal.06}.

\subsection{Physical mechanisms}

\subsubsection{Resuspension term}
The exchange of particles between the current and the substrate includes settling and resuspension contributions.  For grain type $i$ the normalized exchange current, $J_i$, is
\begin{equation}
J_i = J_{si} - J_{ri} = u_{si} ( - \hat{g}_y c_b - \varepsilon_{si} ) ,
\end{equation}
where $J_{si}$ is the settling current of grain type $i$ with a settling velocity, $u_{si}$, such that
\begin{equation}
J_{si} = -u_{si} \hat{g}_y c_b ,
\end{equation}
with $c_b$ being normalized grain volume concentration close to the bottom of the current and $\hat{g}_y = -\cos{\theta}$.  The resuspension current of grains of type $i$ is
\begin{equation}
J_{ri} = u_{si} \varepsilon_{si} ,
\end{equation}
where $\varepsilon_{si}$ is the normalized resuspension volume.  Applying laboratory experiments, \citet{garcia.parker.93} derived the resuspension relation,
\begin{equation}
\label{garcia.parker.eqn}
\varepsilon_{si} = \frac{a}{c_0} \frac{z^5_i}{1+\frac{a}{0.3} z^5_i} f_{ri} ,
\end{equation}
where $f_{ri}$ is a resuspension factor equal to the relative presence of grain type $i$ in the active layer at the substrate surface.  The factor $a$ in Eq. (\ref{garcia.parker.eqn}) for field scale can be increased by a factor of 6 \citep{wright.parker.04}, but can also be reduced by a similar factor due to sediment strength -- the entrainment limiter \citep{kostic.parker.06, fildani.etal.06}.  For simplicity, we use the older value $a=1.3 \times 10^{-7}$ \citep{garcia.parker.93} in our calculation.  The expression for $z_i$ is
\begin{equation}
\label{shear.threshold.eqn}
z_i = \alpha_1 \frac{u_*}{u_{si}} R^{\alpha_2}_{pi} ,
\end{equation}
where $R_{pi}$ is the particle Reynolds number,
\begin{equation}
\label{def.Rpi.eqn}
R_{pi} = \sqrt{R_* g d_i} \frac{d_i}{\nu} ,
\end{equation}
$d_i$ is the diameter of grain type $i$, and $u_* = \tilde{u}_* / u_b$ is the normalized shear velocity at the boundary which can be written in normalized variables as \citep{blanchette.etal.05}
\begin{equation}
\label{norm.shear.eqn}
u_* = \sqrt{ \frac{1}{R_e} \frac{\partial u_x}{\partial y}} .
\end{equation}
We use values for $\alpha_1$ and $\alpha_2$ from experiments by \citet{garcia.parker.93},
\begin{equation}
(\alpha_1,\alpha_2) = \left\{ \begin{array}{ll} (1,0.6) & R_p > 2.36 \\ (0.586,1.23) & R_p \le 2.36 \end{array} \right. .
\end{equation}
In Eq. (\ref{shear.threshold.eqn}), for the geophysical field currents, a slope dependence term $\theta^{0.08}$ of order unity is ignored \citep{kostic.parker.06}.  Equation (\ref{garcia.parker.eqn}) has a high power in $z_i$ and therefore behaves as a threshold relation for the resuspension as a function of ${(u_* / u_{si})}^5$.

\subsubsection{Active layer}
We apply an active layer (AL) in the substrate surface from which resuspension can take place.  Its dimension depends on the resuspension strength \citep{parker.etal.00}.  The AL can have a very large range in depth, from the size of a few grains, in the case of turbidity currents, to the size of the width of the flow, in the case of fluvial flows.  We assume that the flow can mix the particles in the AL, generating a uniform distribution of all grain types in this layer.

The mixing mechanism in the AL can be due to grain traction or small scale topographic variations of the substrate surface.  For example, small scale dunes can accumulate coarse grains in the local minima and fine grains in local maxima.  The AL width would be the long range average distance of these local maxima and minima \citep{parker.etal.00}.  All the grains in the AL are available for resuspension by interaction with the current turbulence.  Resuspension causes a decrease in the upper boundary height of the substrate and for a given AL width, deeper parts of the substrate can now be included in the AL.  Deposition increases the substrate height so that deeper parts of the substrate must now be excluded from the AL.  We include in the computation an AL model capable of handling very small, $L_a \ll 1$, and very large, $L_a\gg 1$, active layers.

In our simulations, we divide the substrate into zones of size $\Delta s$ perpendicular to the substrate surface, where the upper zone may be partially filled.  Typically zones are $\Delta s \approx 0.1$ in normalized units.  The AL can be very large and include many zones, or can be very small and encompass just a fraction of a single zone.  The upper boundary of the substrate is the upper boundary of the AL.  The lower boundary of the AL is obtained by subtracting the AL width, $L_a$, from the upper boundary.  The lower boundary can be in the same zone as the upper boundary or in a much lower zone.  Every time step, we sum over the resuspension and deposition mass, obtain the new upper level of the substrate, and define the AL range.  A mixing process is then applied to the AL to make the grain type distribution uniform from the bottom to the top of the AL.

Armoring happens when fine grains can be resuspended, while coarse are being deposited by the flow.  This will leave an AL made up of only coarse grains which will not be able to be resuspended.  This turns off the resuspension, changing the flow into a purely depositional one.  The result is a dissipating current and reduction in the front velocity.

\subsubsection{Settling velocity}
The settling velocity, $\tilde{u}_{si}$, for grain type $i$ is obtained by using the relationship from \citet{dietrich.82},
\begin{equation}
\tilde{u}_{si} = \sqrt[3]{R_* g \nu W_i} ,
\end{equation}
where
\begin{align}
\log_{10}{W_i} &= -3.76715 + 1.92944 \ A - 0.09815 \ A^2 \nonumber\\ & \quad  - 0.00575 \ A^3 + 0.00056 \ A^4 ,
\end{align}
$A = 2 \log_{10}{R_{pi}}$, and the particle Reynolds number, $R_{pi}$ is defined by Eq. (\ref{def.Rpi.eqn}).  The normalized settling velocities, $u_{si} = \tilde{u}_{si} / u_b$, depends on the input parameters of the particles.  Here, $R_{pi}$ can be identified as the normalized version of the particle diameter, $d_i$.

\subsubsection{Shear factor}
To avoid unrealistic resuspension, a shear factor is introduced.  It is similar to the parameter in other models called the bed resistance coefficient or the bottom drag coefficient, $C_D$ \citep{parker.etal.86}.  The shear factor, $f_{\mathrm{shr}}$, is used to obtain the appropriate shear velocity, $u_*$, to avoid unrealistic resuspension in the simulations.  We include in Eq. (\ref{norm.shear.eqn}) a shear factor, $f_{\mathrm{shr}}$, such that
\begin{equation}
\label{u.eqn}
u_*^2 = \frac{\omega_b}{f_{\mathrm{shr}} R_e} ,
\end{equation}
where $\omega_b$ is the vorticity close to the bottom, $R_e$ is the Reynolds number, and Eq. (\ref{vorticity.eqn}) has been used.  The shear friction force at the bottom of the flow is proportional to $u_*^2$.  Other models such the $\kappa$--$\varepsilon$ turbulence model \citep{felix.01, choi.garcia.02, huang.etal.08} and the depth-averaged model \citep{parker.etal.86} do not have a shear factor.  Instead, they have $C_D$.  We will show that there is a simple relationship between our shear factor, $f_{\mathrm{shr}}$, and $C_D$, so that they are similar parameters.

The equivalent parameter in the other models is defined as
\begin{equation}
\label{cd.def.eqn}
C_D \equiv \left( \frac{u_*}{v_b} \right)^2 ,
\end{equation}
where $v_b$ is the flow velocity at the grid closest to the bottom current.  The relation between $v_b$ and $\omega_b$ is
\begin{equation}
\label{omega.eqn}
\omega_b = \left( \frac{\partial u_x}{\partial y} \right)_b = \frac{v_b}{\Delta_y} ,
\end{equation}
where $\Delta_y$ is the zone height, and $u_x=0$ at the bottom.

In the $\kappa$--$\varepsilon$ turbulence model \citet{felix.01} and \citet{choi.garcia.02} use Eq. (\ref{cd.def.eqn}) to obtain the $u_*$ used in the resuspension relation and approximate $C_D$ by
\begin{equation}
\label{cd.approx.eqn}
C_D = \left( \frac{1}{\kappa} \ln{(E z_b / z_0)} \right)^{-2} ,
\end{equation}
where $\kappa = 0.4$ is the von Karman constant, $E$ is the roughness parameter (which varies between 9 to 30 going from smooth to rough walls), $z_b$ is the height of the lowest grid cell, $z_0$ is the roughness height (for a smooth bottom $z_0=\nu/u_*$), and $\nu$ is the fluid viscosity.  For our case with a scale length of 250 m, 64 zones per unit (that is $z_b=3.9$ m), $E=10$, and $z_0 = 10^{-3}$ m, we get $C_D = 1.4 \times 10^{-3}$.  \citet{felix.01} using Eq. (\ref{cd.approx.eqn}) obtains $C_D = 2.5 \times 10^{-3}$.  \citet{garcia.parker.93} predict that for geophysical currents with Reynolds numbers $R_e \approx 10^{3}$ to $10^{5}$, $C_D \approx 0.1$ to $10^{-3}$.

\citet{parker.etal.86} depth averaged model used three transverse average equations (for height, $h$, velocity, $U$, and concentration, $C$) and used Eq. (\ref{cd.def.eqn}) as a closure condition with $C_D = 4 \times 10^{-3}$.  They also extended the model to four equations, adding an equation for the turbulence energy, $\kappa$.  Assuming that $C_D=\alpha \kappa$ with $\alpha=0.1$, they found that $C_D$ varied in the range of approximately 0.1 to $10^{-3}$.

For our case, we evaluate $C_D$ using Eqs. (\ref{u.eqn})--(\ref{omega.eqn}) and obtain
\begin{equation}
C_D = \frac{1}{f_{\mathrm{shr}} R_e \Delta_y v_b} .
\end{equation}
Substituting our simulation parameters ($R_e = 10^3$, $f_{\mathrm{shr}} = 38$, $\Delta_y = 1/64$, and $v_b = 0.1$) into this equation we get $C_D = 1.7 \times 10^{-2}$.

For our simulations, we evaluated Eq. (\ref{cd.def.eqn}) over a wide range of locations, $x$, and times, $t$.  We found that $C_D$ varies between about 0.1 to $10^{-3}$.  We therefore conclude that $f_{\mathrm{shr}}$ used in our model produces resuspension though a $u_*$ similar to the resuspension obtained in the previous models using $C_D$.

\subsection{Numerical approach}
\label{num.approach}
The numerical methods used to solve the current Eqs. (\ref{lie.derivative.vorticity})--(\ref{lie.derivative.c}) are based on \citet{lele.92, hartel.etal.00, blanchette.etal.05}.  We perform a Fourier transform on $\psi$ in the $x$ direction and use a sixth-order finite difference scheme for the derivatives in the $y$-direction, except near the boundaries where the the derivatives are accurate to third order.  A third-order Runge-Kutta integrator is used to propagate the solution in time.  A finite difference time integrator is applied to Eq. (\ref{exner.eqn}) to update the substrate particle budget, and an AL scheme is applied in the determination of the balance between erosion and deposition.  An adaptive time step is used which satisfies the Courant-Friedrichs-Levy condition while minimizing computation time.  A typical time step is $\Delta t \approx 0.01$.  For a typical length scale $L_0 = 250$ m and buoyancy velocity $u_b = 5$ m/s, we get a time scale of $t_0 = L_0 / u_b = 50$ s.

Typically, the fluid equations are solved over a rectangular domain ($-4\le x \le 23$ and $0 \le y \le 3$) divided into 513 and 193 grid cells, respectively.  An additional rectangular grid is used for the substrate at the same $x$ locations and over a range, in the perpendicular direction, of $0<s<20$, where the distance is scaled by $s_0 = L_0 c_ 0 / (1-\lambda_p)$ with a porosity of $\lambda_p = 0.3$.  The substrate is divided into 513 and 601 grid cells in the $x$ and $s$ directions, respectively.  An AL is applied of height $L_a=0.02$.  Changing $L_a$ by a factor of 2 has a small effect on the results.

A lock release is simulated where the fluid is initially located at rest in a typical range of $-4 \le x \le 0$ and $0 \le y \le 1.5$, with a fluid height of $y=3$.  It was shown by \citet{blanchette.etal.05} that the effect of the upper fluid boundary can be neglected if it is at least twice the height of the lock release, that is it can be considered a deepwater case.  For numerical stability the initial lock particle concentration profile and the substrate bottom topography is smoothed over a few grid points (typically 6).  The typical initial volume concentration is $c_0 = 0.8\%$.

The value of Reynolds numbers, $R_e = u_b L_0 / \nu$, for geological turbidity currents with $u_b$ in the range of 1 m/s to 5 m/s, scale lengths, $L_0$, in the range of 1 m to 250 m, and a water viscosity of $\nu = 10^{-6} \ \mathrm{m}^2/\mathrm{s}$, are in the range $10^6$ to $10^{10}$.  These Reynolds numbers are well beyond the reach of numerical simulation.  As $R_e$ increases, smaller scales must be resolved, which also implies smaller time steps as well as more grid cells.  However, as shown by \citet{blanchette.etal.05} there is little change to the eddy structure as $R_e$ increases from $10^3$ to $10^4$.  Therefore we have used $R_e = 10^3$ in this work.  Although we capture only the large scale behavior of the current, neglecting the smaller scales should not change the result at these larger scales.

To avoid unphysical resuspension, we use a shear factor of $f_{\mathrm{shr}}=38$.  The eroded particles are spread uniformly in a region close to the substrate, typically over a thickness of 0.15.  Changing the spreading range by 20\% only has a small effect on the results.  When the resuspension is high, the particles injected over this range are rapidly transported further by the flow to distances much greater than the initial injection range.  This leads to the small sensitivity to the initial injection range.  In contrast, depth-averaged models the injected particles are spread over the transverse layer.

Five types of grains are simulated with diameters that range from 300 $\mu$m to 1000 $\mu$m.  The number of flows simulated are typically 120.  A typical runtime on an 8 core (dual quad core Xeon E5462, 2.68 GHz) machine is 20 hours.  The program is restartable.

\subsection{Transverse average current variables}
To study the current structure and compare it to previous work, we depth average the transverse current profiles in the $y$-direction as a function of $x$.  In the appendix of the paper by \citet{parker.etal.86}, they write the depth-averaged variables for the velocity, $U$, height, $h$, and concentration, $C$, in terms of the local velocity, $u_x$ as
\begin{equation}
\label{average.u.eqn}
U \ h = \int_{0}^{y_l} u_x \, dy = a_1 ,
\end{equation}
\begin{equation}
U^2 \ h = \int_{0}^{y_l} u^2_x \, dy = a_2 ,
\end{equation}
and
\begin{equation}
\label{average.c.eqn}
U \ C \ h = \int_{0}^{y_l} u_x \, c \, dy = a_3 ,
\end{equation}
where $y$ is the transverse coordinate, $u_x$ is the velocity in the longitudinal $x$-direction, and $c$ is the normalized concentration.  Here, $y_l$ defines the range in the $y$-direction of appreciable concentration in the current, $c>c_l$ \citep{middleton.93}.  We use a value of $c_l = 3/4$, relative to the initial concentration of 1.  The transverse layer average values for $U$, $h$, $C$ obtained from Eqs. (\ref{average.u.eqn})--(\ref{average.c.eqn}) are
\begin{equation}
\label{ut.eqn}
U = \frac{a_2}{a_1} ,
\end{equation}
\begin{equation}
h = \frac{a^2_1}{a_2} ,
\end{equation}
and
\begin{equation}
\label{ct.eqn}
C = \frac{a_3}{a_1} .
\end{equation}
Averaging the simulation profiles of $u_x$ with Eqs. (\ref{average.u.eqn})--(\ref{average.c.eqn}), the transverse average variables $U$, $h$, and $C$ are obtained by using Eqs. (\ref{ut.eqn})--(\ref{ct.eqn}).

The depth-average variables are used to calculate the local Richardson number,
\begin{equation}
Ri = \frac{1}{F^2_r} = \frac{R_{*} g C h}{U^2} ,
\end{equation}
where $F_r$ is the local Froude number.  We define $F_2$ as
\begin{equation}
F_2 \equiv \frac{1}{Ri}-1 .
\end{equation}
The square of the Froude number is proportional to $U^2 / h$ and indicates the ratio of the kinetic energy of the flow to the potential energy of the fluid.  For $Ri<1$, $F_r>1$, or $F_2>0$ the local current is supercritical, that is the kinetic energy of the flow is greater than the potential energy of the fluid.  It is usually assumed \citep{parker.etal.86} that a supercritical flow is predominately erosional and that a subcritical flow, $F_2<0$, is predominately depositional.

Eventhough the depth-averaged variables only show the characteristics of the envelope of the current, we will find that the eddy structure has an imprint on the average velocity, $U$, and average height, $h$.  There are periodic structures on these variables correlated with both the sediment waves and eddy structure.  A convergence of the flow toward the substrate reduces $h$ and increases $U$ causing a peak in $F_2$.  This peak correlates with a peak in the shear velocity, $u_*$, and with a resulting increase in resuspension.  In the next section we will present the depth average variables $U(x)$, $C(x)$, and $h(x)$, obtained from the detailed current profile, as functions of the location $x$.  We will also present the local Froude number dependance as $F_2(x)$ (remember that $F_2>0$ indicates supercritical flow locally and $F_2<0$ indicates locally subcritical flow), and the local $u^5_*(x)$ (indicating the local shear velocity dependance of the resuspension, see Eqs. (\ref{garcia.parker.eqn}) and (\ref{shear.threshold.eqn})).

\section{Simulation results for sediment wave generation}
\label{simulation.results}

\subsection{Effect of an obstacle}
\begin{figure}
\noindent\includegraphics[width=20pc]{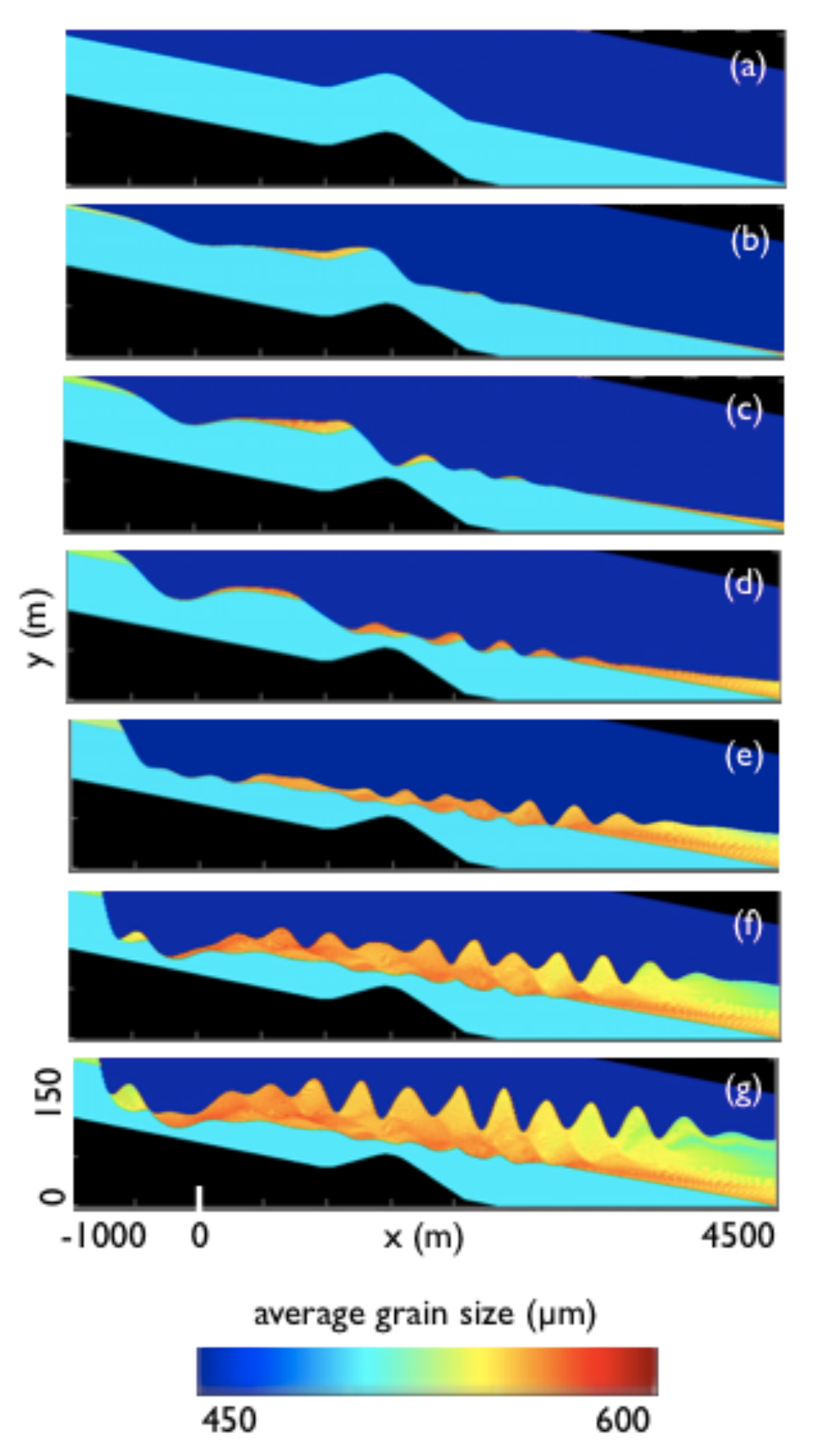}
\caption{\label{fig1} Development of sediment waves on the substrate.  The $x$-$y$ image is colored according to the average grain diameter.  The color bar is a rainbow, starting from 450 $\mu$m at blue and ending at 600 $\mu$m at red.  The profiles are shown after:  (a) 1, (b) 5, (c) 10, (d) 20, (e) 40, (f) 80, and (g) 120 flows.  The initial slope is $\theta_0 = 1.5^{\circ}$, and the flows include 5 types of grains with diameters that range from 300 $\mu$m to 700 $\mu$m ($R_{pi}=20$ to 71).  Initially there is an obstacle between $x$ values of 1000 m and 2000 m with a peak at 1500 m (in normalized units: 4, 8, and 6).  The upstream slope of the obstacle is $-2^{\circ}$, and the downstream slope is $5^{\circ}$.  The lock is between -1000 m and 0 m ($-4<x<0$) with an initial height of 375 m ($H=1.5$) and particle concentration of $c_0=0.8\%$.} 
\end{figure}

We simulated multiple lock release flows down a 2D ``channel'' in the $x$-$y$ plane of dimension $-4<x<23$ and slope $\theta_0 = 1.5^{\circ}$, where the scale length for $x$ and $y$ is $L_0=250$ m.  The flow and the substrate initially include 5 types of grains equally distributed with diameters of $\{d_i\}=\{300,400,500,600,700\} \mu$m.  This corresponds to particle Reynolds numbers, $R_{pi}$, that range from 20 to 71.  The suspension in the lock is located in the area where $-4<x<0$,  $0<y<H$, and $H=1.5$.  The water boundary is at $y=3$, which is large enough to cause little coupling of the flow to the water boundary -- a deepwater flow.  The initial particle concentration in the lock is $c_0=0.8\%$.  An obstacle of triangular shape with rounded corners is located along the channel at $\{x_i\}=\{4,6,8\}=\{\mathrm{start},\mathrm{top},\mathrm{end}\}$ with an angle of $-2^{\circ}$ on the upstream side and $5^{\circ}$ on the downstream side.  The current is absorbed at the end of the channel, in the range of $x=20$ to 23.  The initial substrate and obstacle structure is presented (in real units) in Fig. \ref{fig1}a.

Figure \ref{fig1} presents the substrate structure and the development of the sediment wave along the channel in the $x$-$y$ plane for up to 120 sequential flows.  Each flow has been completed before the next is started.  The substrate is colored according to the average grain diameter over the range of 450 $\mu$m to 600 $\mu$m.  Considering the particle diameter distribution in the substrate, by examining its width or variance, we find similar behavior.  For the fifth flow, $f=5$, there is deposition before the obstacle crest and erosion after the crest.  The extra erosion downstream generates the next break (increase) in slope and starts the next crest in the downstream direction.  For $f=10$ and $f=20$ a train of breaks (increase) in slope develop seeding the SW structure.  Every SW crest moves upstream due the current deposition on the upstream side of the crest and the erosion on the downstream side.  By $f=40$ a well developed SW train is formed, propagating downstream by the seeding of new breaks in slope, and migrating upstream by the structured erosion and deposition.  By $f=80$ and $f=120$ the upstream SW are effected by erosion close to the lock boundary.  The downstream part of the SW starts to be affected by the change in slope due to current reflection and deposition at the right boundary.  Increasing the channel length extends the range of the SW downstream, but does not effect the general structure of the flows that we will analyze.  In the figures presenting the development of SW in the substrate, as in Fig. (1), we consider the color map of the average grain size diameter. 

\begin{figure}
\noindent\includegraphics[width=20pc]{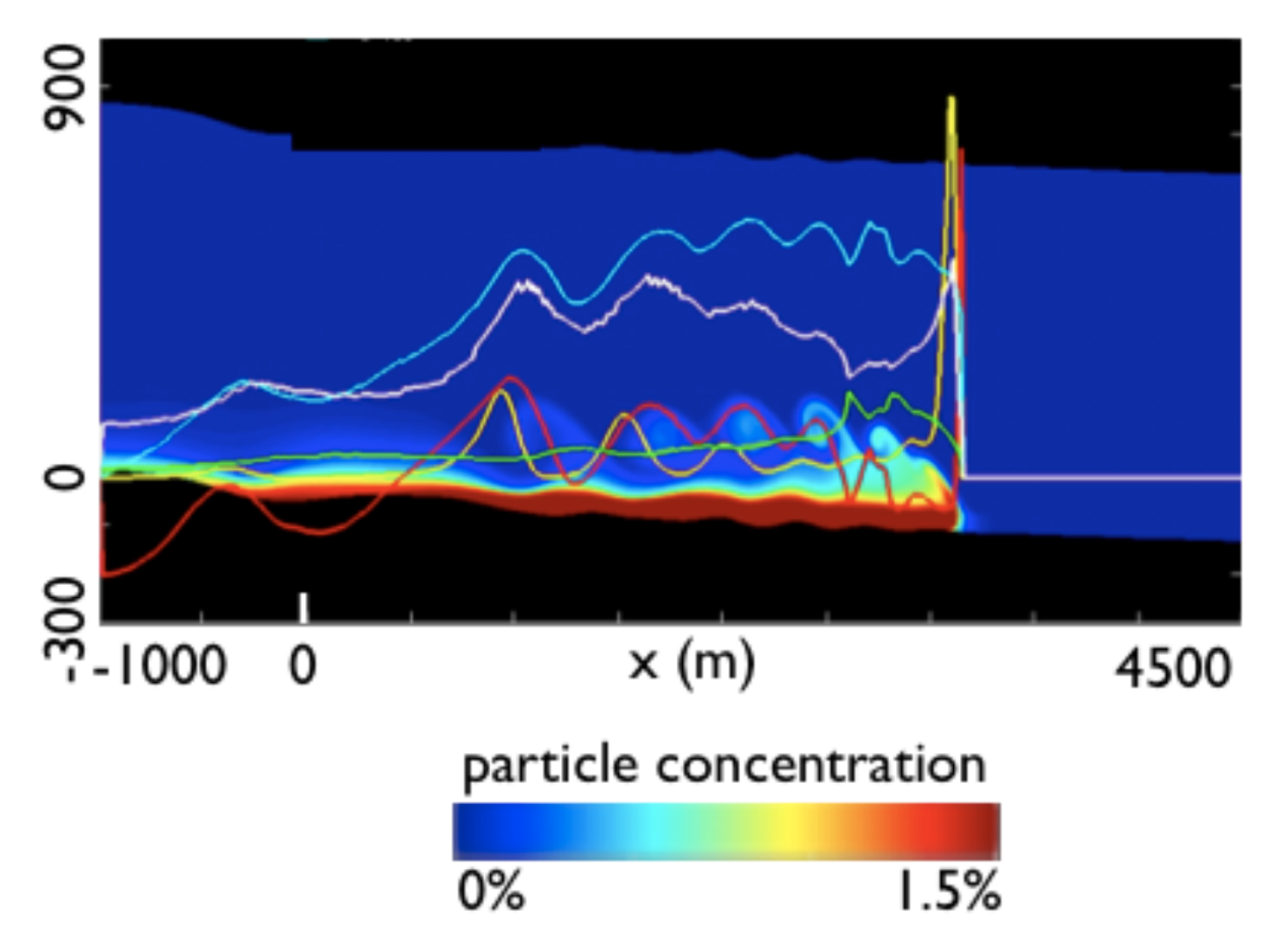}
\caption{\label{fig2} Particle volume concentration of the flow in the $x$-$y$ plane at the normalized time, $t=8$, for flow, $f=20$ (time scale is $t_0=L_0/u_b=46$ s, buoyancy velocity is $u_b=\sqrt{gR_{*}c_0 L_0}=5.42$ m/s, particle density change relative to water is $R=1.5$, and initial particle concentration is $c_0=0.8\%$).  This corresponds to Fig. \ref{fig1}c.  The color bar is a rainbow, starting from 0 at blue and ending at 1.5\% at red.  Also shown are the depth averaged current variables:  (blue) velocity $U \times 100$ in m/s, (white) concentration $C \times 2 \times 10^4$, (green) current height $h$ in m, (red) change in Froude number $F_2 \times 200$, and (yellow) shear velocity term in the resuspension $V_{\mathrm{shr}}^5 \times 5 \times 10^6$ in (m/s)$^5$.  All quantities plotted have SI dimensions.} 
\end{figure}

Figure \ref{fig2} presents contours of the current's particle concentration in the $x$-$y$ plane at the normalized time, $t=8$, for flow, $f=20$.  The current head has already passed over the obstacle.  The time scale is $t_0=L_0/u_b=46$ s, the length scale is $L_0= 250$ m, the buoyancy velocity is $u_b=\sqrt{gR_{*}c_0 L_0}=5.42$ m/s, $g=9.81$ m/s$^2$, and the particle density change is $R_{*}=1.5$.  The image colors are the particle volume concentration in the range of 0 to 1.5\%.  Also shown in Figure \ref{fig2} are the transverse average variables as a function of $x$: the current velocity $U(x)$ in blue, the concentration $C(x)$ in white, the current height $h(x)$ in green, the change in the Froude number $F_2(x)=F_r^2-1=1/Ri-1$ in red, and the shear velocity term in the resuspension expression $V_{\mathrm{shr}}^5$ in yellow.  Note that SW periodicity in the substrate is coupled into the current and appears as eddies in the current and as a periodicity in the transversely averaged variables.  Characteristic values for the flow are $U \approx 4.5$ m/s, $C \approx 1.5\%$ (twice its initial value), and $h \approx 60$ m.  The Froude number is greater than 1 for a large part of the flow (supercritical) and the flow is highly erosive.  Note that the amount erosion is not well correlated with the degree of supercriticality.  There is also an exponential growth in the erosion as one goes from the head to the tail of the flow, while the change is the Froude number is relatively constant.  The wavelength of the SW is consistent with the \citet{normark.etal.80} relation, $\lambda=2 \pi h$, where $\lambda=380$ m for $h=60$ m.

\begin{figure}
\noindent\includegraphics[width=20pc]{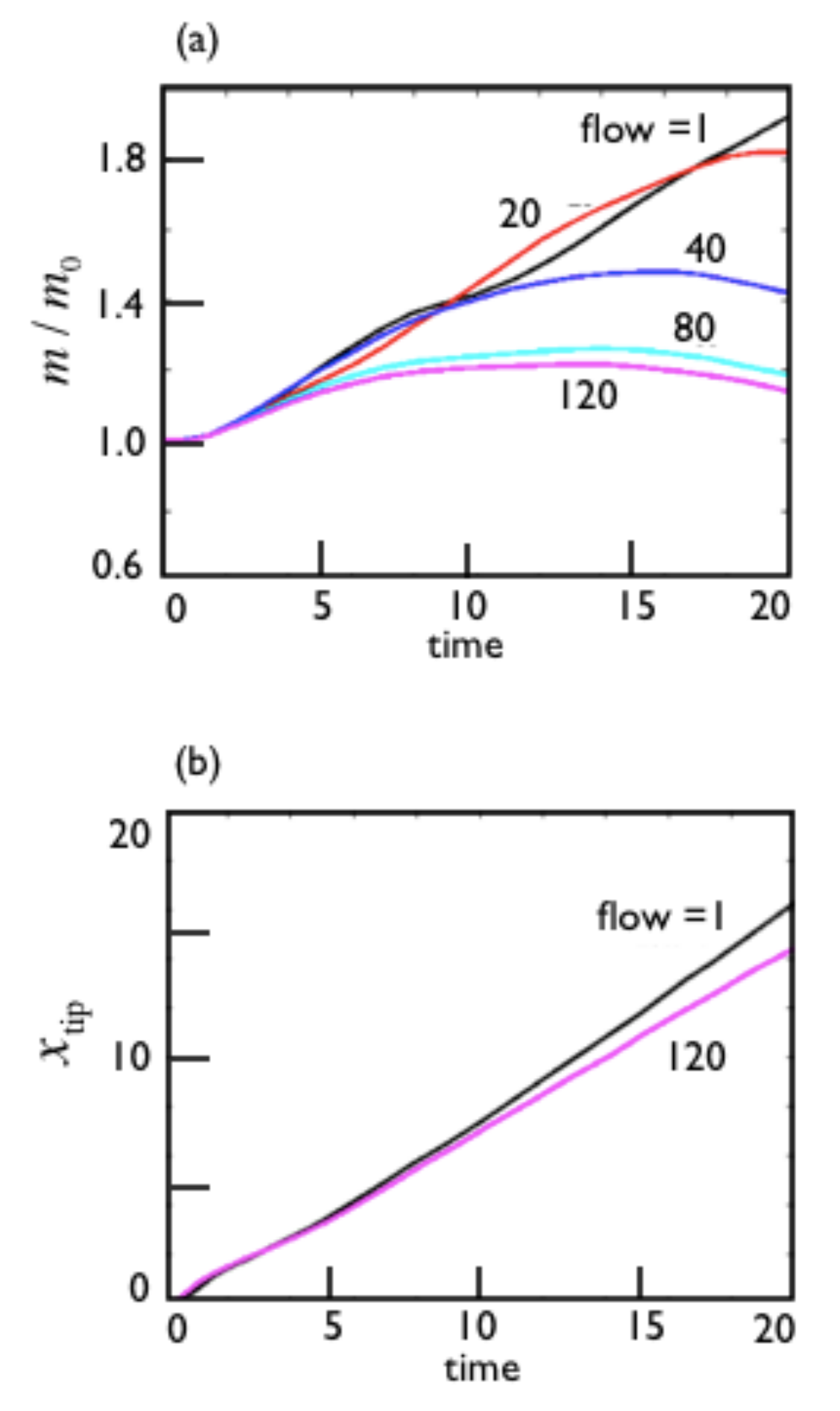}
\caption{\label{fig3} (a) normalized suspended mass in flow as a function of time, $m(t)/m_0$, for flows:  (black) 1, (red) 20, (blue) 40, (cyan) 80, and (magenta) 120.  Time is plotted in normalized units (the scale for time is $t_0=46$ s).  (b) front position as a function of time, $x_{\mathrm{tip}}(t)$, for flows:  (black) 1 and (magenta) 120.  The front velocity is reduced from 0.90 for flow 1, to 0.73 for flow 120.  In dimensional units, these are velocities of 4.9 m/s and 3.9 m/s ($u_b=5.42$ m/s).} 
\end{figure}

Figure \ref{fig3} shows the total mass in the flow, $m(t)$, and its front position, $x_{\mathrm{tip}}(t)$, as a function of time for flows $\{f\}=\{1,20,40,80,120\}$.  For flows 1 and 20, there is an increase in the total mass, because of resuspension, by almost a factor of 2.  The current asymptotes to a speed of about 4.5 m/s.  The resuspension maintains the current motion and redistributes the substrate mass to form the growing SW.  For later flows (40, 80, and 120), the substrate slope is reduced by deposition at the end of the channel.  Consequently, the resuspension and the growth of the SW are reduced, keeping the mass in the flow constant and changing only the sediment wave structure.

\begin{figure}
\noindent\includegraphics[width=20pc]{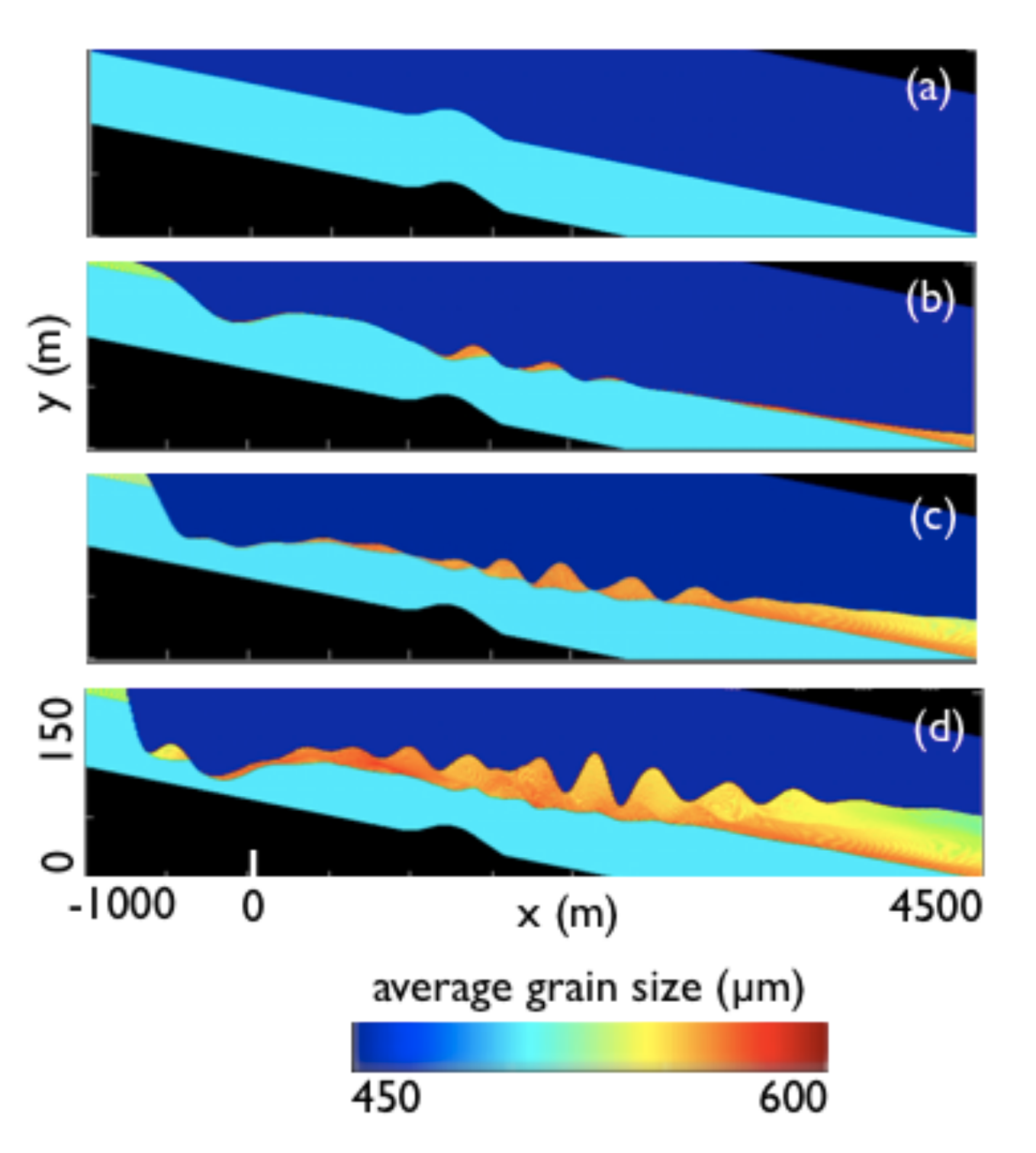}
\caption{\label{fig4} Development of sediment waves on the substrate with a reduced size obstacle (by a factor of 2).  Initially there is an obstacle between $x$ values of 1000 m and 1500 m with a peak at 1250 m (in normalized units: 4, 6, and 5).  The upstream slope of the obstacle is $-2^{\circ}$, and the downstream slope is $5^{\circ}$.  The $x$-$y$ image is colored according to the average grain diameter.  The color bar is a rainbow, starting from 450 $\mu$m at blue and ending at 600 $\mu$m at red.  The profiles are shown after:  (a) 1, (b) 10, (c) 40, and (d) 80 flows.  Other than the size of the obstacle, all parameter are identical to the simulations shown in Figs. \ref{fig1} to \ref{fig3}.} 
\end{figure}

Figure \ref{fig4} shows the result of reducing the obstacle height and width by a factor of 2, located at $\{x_i\}=\{4,5,6\}$ with angles $\{\theta_i\}=\{-2^{\circ},5^{\circ}\}$.  All other parameters are the same as the previous simulation.  Reducing the obstacle size has only a small effect on the growing SW.  Comparing Fig. \ref{fig4} to Fig. \ref{fig1} for flow 80, we find a very similar SW development.  The obstacle's function is to trigger the probable growing wavelength.  By flow 10, the obstacle is eroded leaving the system to develop SW independent of the initial condition.

\subsection{Influence of lock height}
\begin{figure}
\noindent\includegraphics[width=20pc]{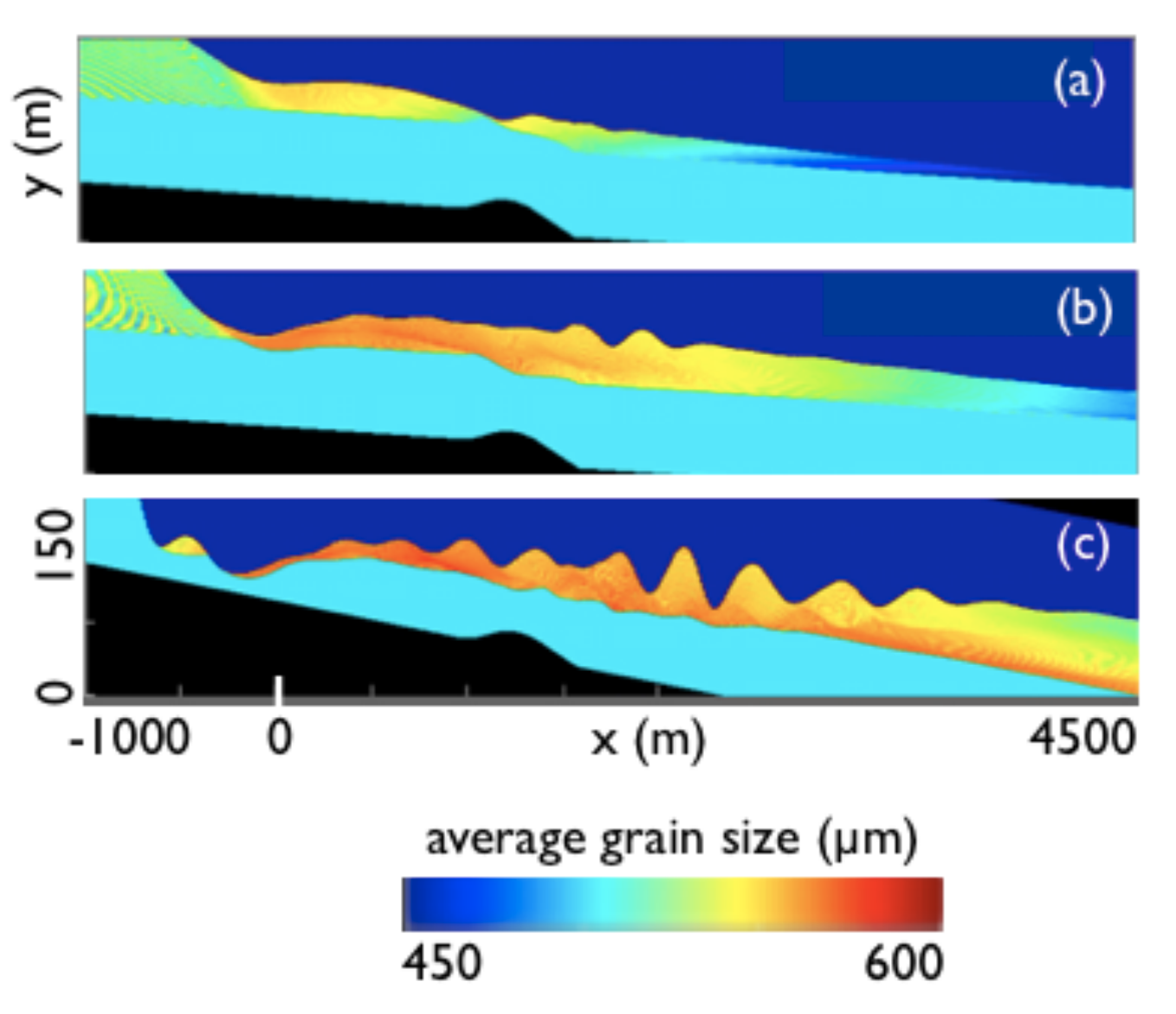}
\caption{\label{fig5} Development of SW on a substrate after 80 flows, to study the effect of the initial height, $H$.   The $x$-$y$ image is colored according to the average grain diameter.  The initial obstacle has the reduced height shown in Fig. \ref{fig4}.  The color bar is a rainbow, starting from 450 $\mu$m at blue and ending at 600 $\mu$m at red.  The profiles are shown for:  (a) ``no SW'', $H=0.5$, $\theta_0=0.5^{\circ}$;  (b) ``SW buildup'', $H=1.0$, $\theta_0=0.5^{\circ}$; and (c) ``SW growth'', $H=1.5$, $\theta_0=1.5^{\circ}$.} 
\end{figure}

\begin{figure}
\noindent\includegraphics[width=20pc]{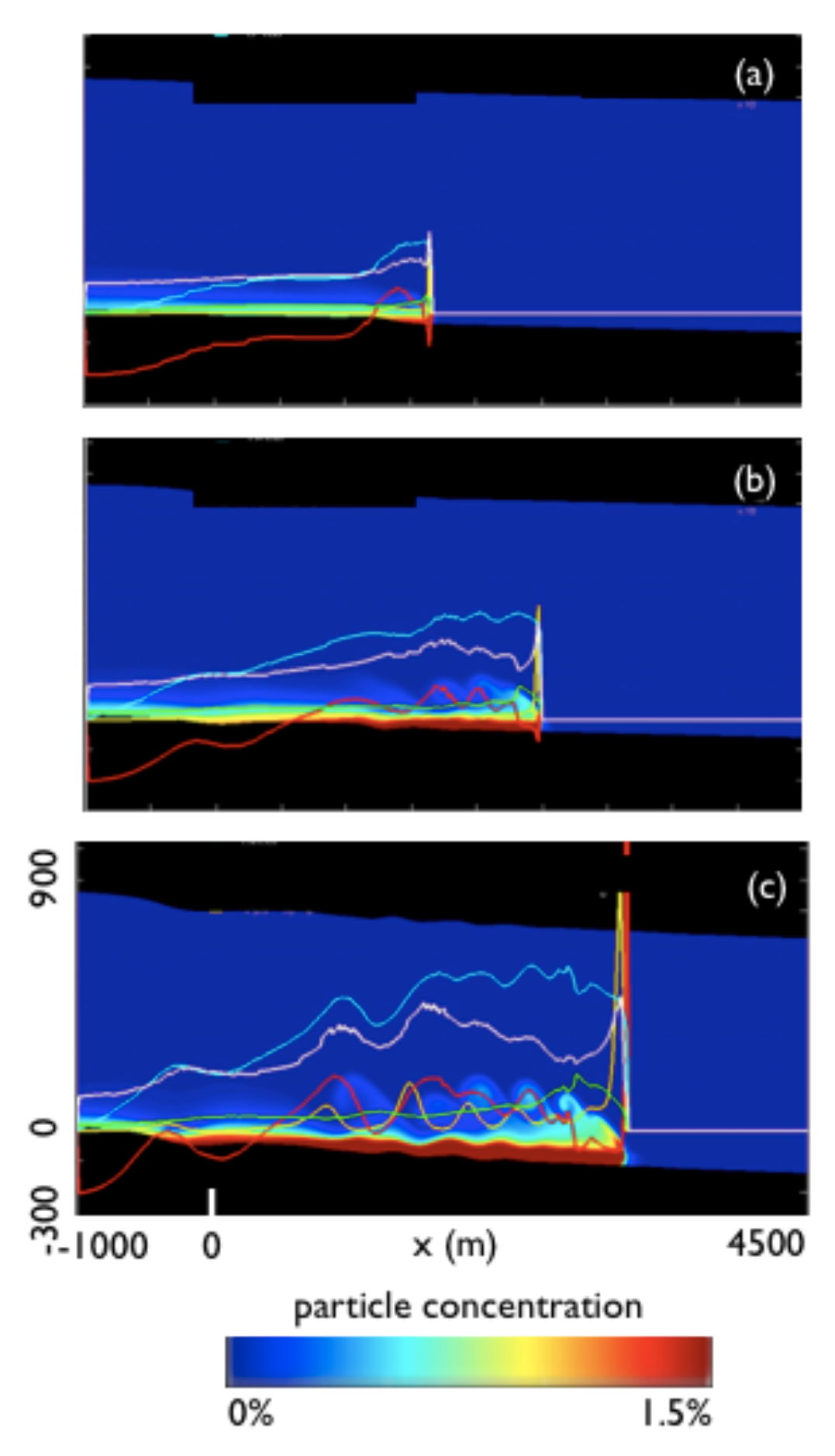}
\caption{\label{fig6} Particle volume concentration of the flow in the $x$-$y$ plane at the normalized time, $t=8$, for flow, $f=15$, to study the effect of the initial height, $H$.  This corresponds to simulations of Fig. \ref{fig5}.  The color bar is a rainbow, starting from 0 at blue and ending at 1.5\% at red.  Also shown are the depth averaged current variables:  (blue) velocity $U \times 100$ in m/s, (white) concentration $C \times 2 \times 10^4$, (green) current height $h$ in m, (red) change in Froude number $F_2 \times 200$, and (yellow) shear velocity term in the resuspension $V_{\mathrm{shr}}^5 \times 5 \times 10^6$ in (m/s)$^5$.  All quantities plotted have SI dimensions.  The profiles are shown for:  (a) ``no SW'', $H=0.5$, $\theta=0.5^{\circ}$;  (b) ``SW buildup'', $H=1.0$, $\theta=0.5^{\circ}$; and (c) ``SW growth'', $H=1.5$, $\theta=1.5^{\circ}$.} 
\end{figure}

\begin{figure}
\noindent\includegraphics[width=20pc]{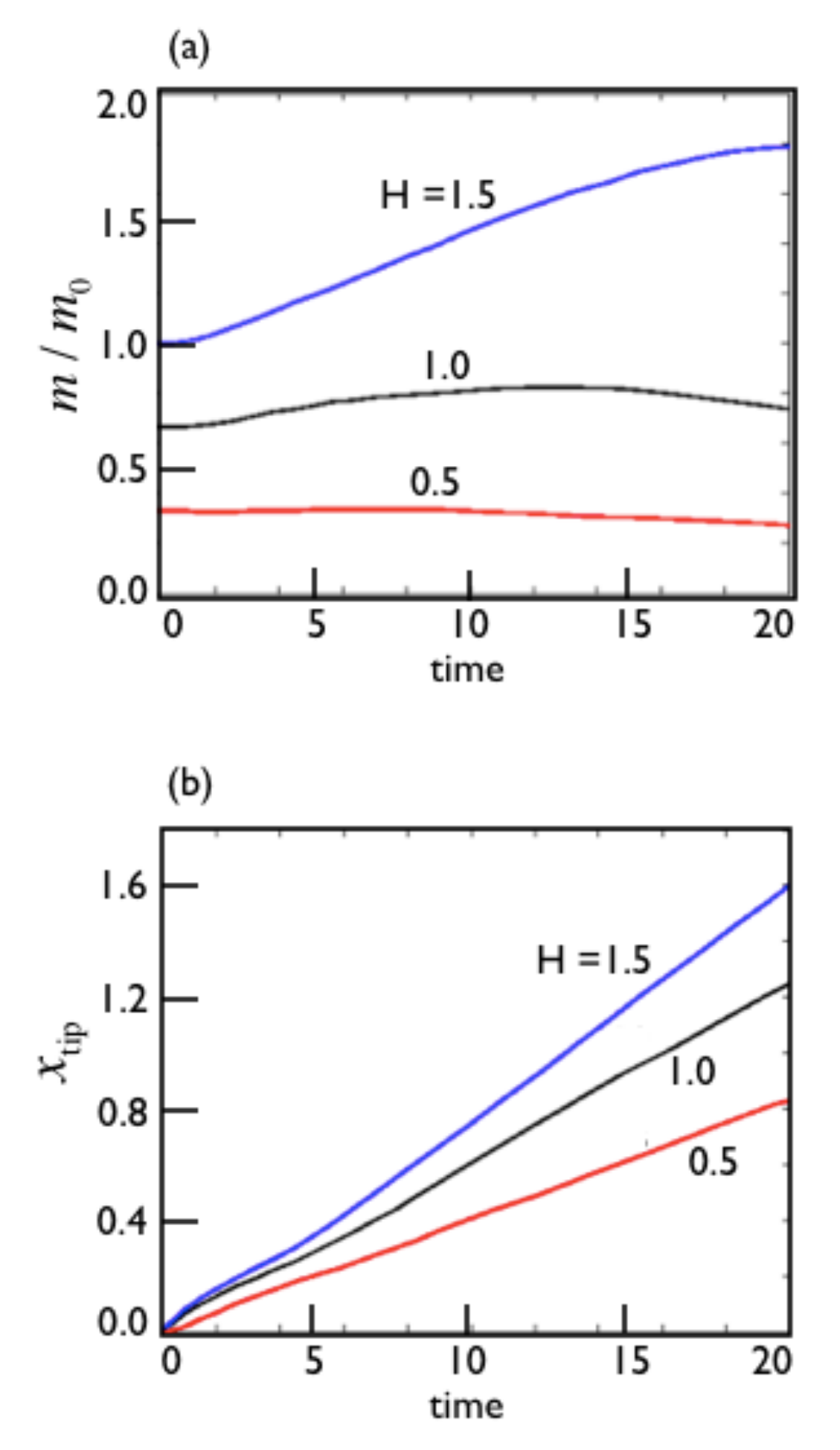}
\caption{\label{fig7}  To study the effect of the initial height, $H$. (a) normalized suspended mass in flow as a function of time, $m(t)/m_0$, for the same simulations as Fig. \ref{fig5} and \ref{fig6}:  (red) ``no SW'', $H=0.5$, $\theta=0.5^{\circ}$;  (black) ``SW buildup'', $H=1.0$, $\theta=0.5^{\circ}$; and (blue) ``SW growth'', $H=1.5$, $\theta=1.5^{\circ}$.  Time is plotted in normalized units (the scale for time is $t_0=46$ s).  (b) front position as a function of time, $x_{\mathrm{tip}}(t)$.  The front velocity is reduced to 0.42 for ``no SW'', maintained at 0.66 for ``SW buildup'', and increased to 0.87 for ``SW growth''.  In dimensional units, these are velocities of 2.3 m/s, 3.6 m/s, and 4.7 m/s.} 
\end{figure}

We now present a series of systematic parameter studies over the next three subsections of the paper.  We start with examining the influence of lock height, $H$, on the character of the flow and SW formation.  The lock height is directly related to the size of the flow.  Three characteristic lock heights (0.5, 1.0, and 1.5) are simulated for the reduced obstacle system displayed in Fig. \ref{fig4}.  The results are shown in Figs. \ref{fig5} to \ref{fig7}.  The slope angle was also changed for each of the cases to 0.5$^{\circ}$, 0.5$^{\circ}$ and 1.5$^{\circ}$, respectively.  This was done to access the three different phases of SW development.

For the first case (Fig. \ref{fig5}a), there was no sediment wave formation.  We call this the ``no SW'' phase.  It is characterized by a final uniform slope topography.  As more flows are deposited the obstacle is removed from the topography.  Further characteristics of this phase can be seen in Figs. \ref{fig6}a and \ref{fig7}.  Figure \ref{fig6} shows the profile of the particle concentration in flow and the depth averaged current variables in the same manner as Fig. \ref{fig2}.  The time evolution of the suspended mass in the flow, $m(t)$, and the front position, $x_{\mathrm{tip}}(t)$, are shown in Fig. \ref{fig7} in the same manner as Fig. \ref{fig3}.  Note the simple structure to the flow in Fig. \ref{fig6}a.  The flow is divided into the head with and elevated velocity and concentration.  It is modestly supercritical as evidenced by $F_2$.  The head is followed by a subcritical body.  There is no appreciable erosion as evidenced by the small values of $V_{\mathrm{shr}}^5$.  There is little structure within these two parts of the flow.  Figure \ref{fig7} shows a monotonically decreasing mass and a reduced front velocity of 0.42 in normalized units and 2.3 m/s in dimensional units.  An important thing to note about this phase is that the initial substrate is at no point steep enough in slope to support self sustainment according to the criteria presented in \citet{blanchette.etal.05}.  This criteria gives the critical angle, $\theta_c$, for self sustainment as a function of $c_0$, $H$, and $d$.  The characteristics of the deposited beds shown in Fig. \ref{fig5}a, are quite simple.  Eventhough there have been many flows there appears to be one massive bed that becomes gradually more fine grained downslope and gradually more coarse grained going from the bottom to the top of this massive bed.

The second phase is demonstrated in Figs. \ref{fig5}b, \ref{fig6}b, and \ref{fig7}.  We call this phase ``SW buildup''.  This phase is characterized by the obstacle being reorganized by the early flows into a stable self-consistent profile that neither grows or decays with additional flows.  It should be noted that the initial substrate profile is only steep enough on the downstream side of the obstacle to support self sustainment.  We recognize that this profile is maintained on the boundary of SW growth where the resuspension leading to growth is balanced by the deposition favoring decay.  Because of this and the invariant profile that we call this phase soliton like.  In addition, it is very similar to the buildup mode in a laser.  In a buildup mode, random perturbations in the laser cavity are self organized into a persistent organized mode in the laser cavity.  This is the reason for the name of this phase.  Further characteristics of this phase are shown in Fig. \ref{fig6}b.  The flow is now modestly supercritical over most of its evolution as evidenced by the $F_2$ profile.  It also shows structure in the velocity, concentration, and especially $F_2$ that is synchronised to the SW structure.  There is still no appreciable erosion as evidenced by $V_{\mathrm{shr}}^5$.  Figure \ref{fig7} shows that $m(t)$ has a maximum and remains near the initial mass.  The front velocity of 0.66, 3.6 m/s in dimensional units, is not elevated or reduced.   The characteristics of the deposited beds shown in Fig. \ref{fig5}b, display a bit more structure than the previous phase.  There still do not appear to be distinct beds associated with each flow.  Instead there is a massive bed with gradually changing characteristics.  It becomes more fine grained downslope.  Vertically it shows more character that the previous phase.  It gradually oscillates from bottom to top.  The resulting profile has stripes of coarse grained deposits dipping down in the downslope direction.

The third phase is demonstrated in Figs. \ref{fig5}c, \ref{fig6}c, and \ref{fig7}.  We call this phase ``SW growth''.  This phase is characterized by a SW that initially grows exponentially.  It is seeded from the obstacle generating a sequence of SW crests in the downstream direction.  The wave then migrates slowly upstream.  The obstacle is removed from the substrate by the early flows and the subsequent evolution has no memory of the initial obstacle.  It should be noted that the initial substrate profile is always steep enough to support self sustainment.  Further characteristics of this phase are shown in Fig. \ref{fig6}c.  The flow is significantly supercritical over the body and is marginally supercritical near the head as evidenced by the $F_2$ profile.  It shows structure in the velocity, concentration, $F_2$, and erosion parameter, $V_{\mathrm{shr}}^5$ that is synchronised to the SW structure.  A distinguishing characteristic of this phase is the appreciable erosion evidenced by the $V_{\mathrm{shr}}^5$ profile.  It also shows a exponentially growing (from head to tail) wave structure that is synchronised to the SW structure.  Figure \ref{fig7} shows that $m(t)$ is monotonically increasing and approaches an asymptote that is about twice the initial mass.  The front velocity of  0.87, 4.7 m/s in dimensional units, is elevated.  The characteristics of the deposited beds shown in Fig. \ref{fig5}c, are quite complex.  There are distinct beds for each flow.  There is an overprint of a complex structure as the SW migrate upstream and erode into the substrate.

\begin{figure}
\noindent\includegraphics[width=20pc]{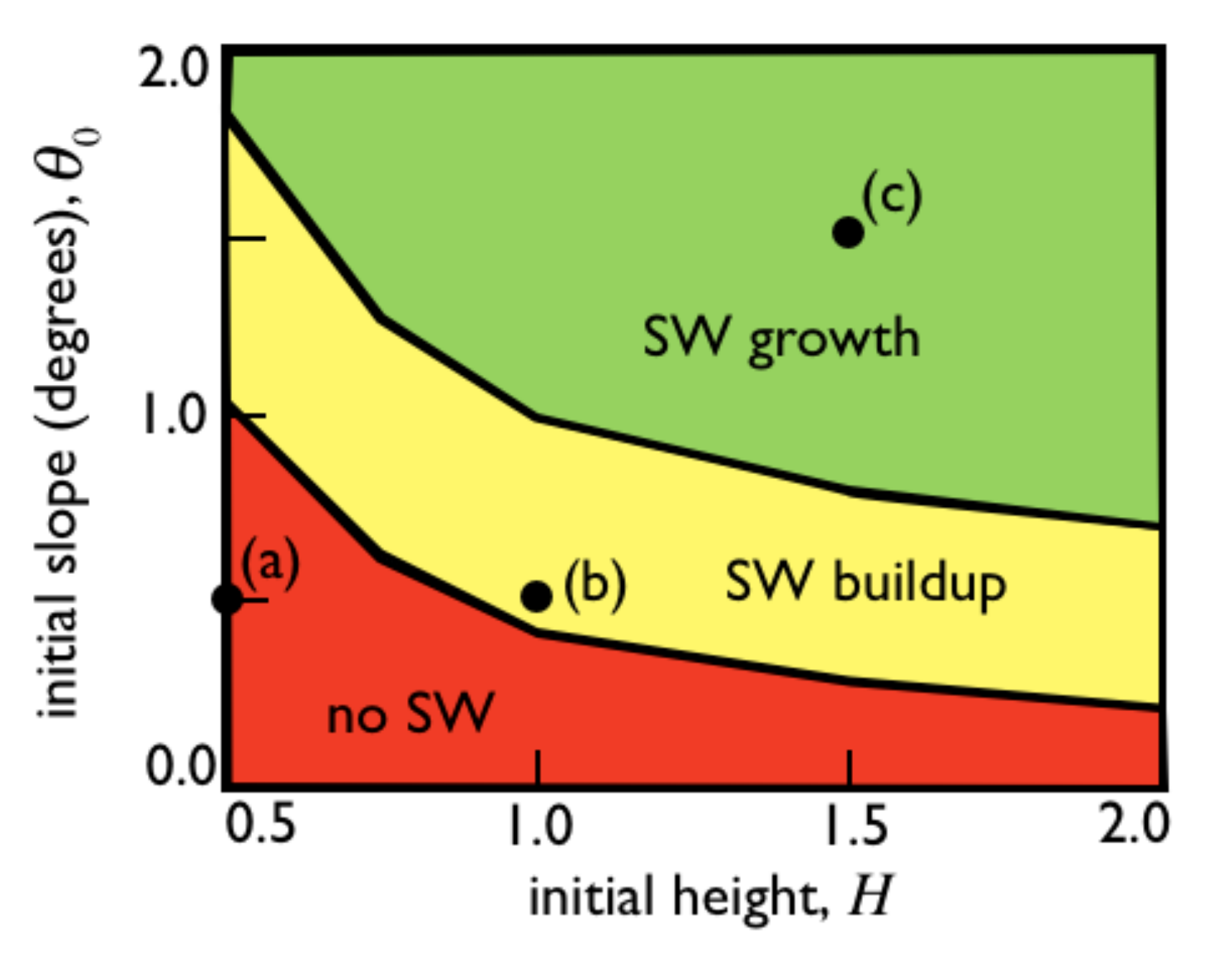}
\caption{\label{fig8} Phase diagram for SW in the $H$-$\theta_0$ plane, where $H$ is the initial lock height in normalized units and $\theta_0$ is the initial slope of the substrate.  The three regions are identified according to the phase of the SW:  (red) ``no SW'', (yellow) ``SW buildup'', and (green) ``SW growth''.  The points simulated in Figs. \ref{fig5} to \ref{fig7} are plotted as black dots and labeled (a, b, and c) consistent with those previous figures.} 
\end{figure}

The picture of these phases is completed by a much larger set of simulations that were done over a large range of lock height, $H$, and slope angle, $\theta_0$.  For each of the simulations the flow was classified by what phase of SW developed (``no SW'', ``SW buildup'', or ``SW growth'').  The results are displayed in Fig. \ref{fig8}.  This figure divides the $H$-$\theta_0$ plane into three regions depending on the phase of the SW.  The three exemplars shown in the previous three simulations are indicated as the black points on this figure.  As $H$ increases, $\theta_0$ can be reduced and still maintain the SW phase.  You can now see why we had to increase the angle, as well as the initial flow height, to have the third case be in the ``SW growth'' phase.  This figure is a cut through the phase space at constant initial particle concentration, $c_0$, and particle size, $\{d_i\}$.  The behavior of the phase diagram with these remaining two variables will be explored in the next two subsections.

\subsection{Effect of particle concentration}
\begin{figure}
\noindent\includegraphics[width=20pc]{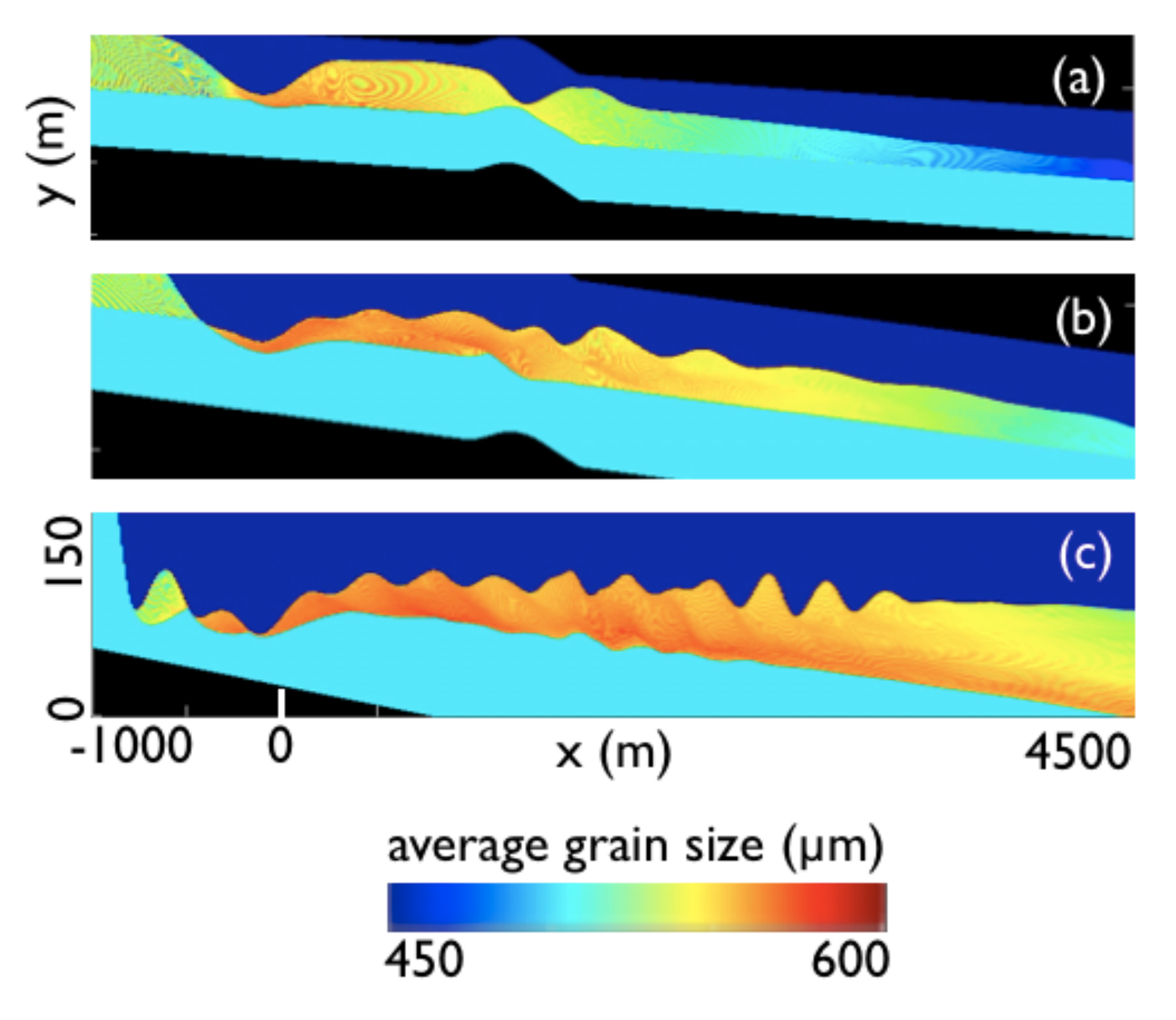}
\caption{\label{fig9} Development of SW on a substrate after 80 flows, to study the effect of the particle concentration, $c_0$.   The $x$-$y$ image is colored according to the average grain diameter.  The initial obstacle has the reduced height shown in Fig. \ref{fig4}.  The color bar is a rainbow, starting from 450 $\mu$m at blue and ending at 600 $\mu$m at red.  The profiles are shown for:  (a) ``no SW'', $c_0=0.4\%$, $\theta_0=0.5^{\circ}$;  (b) ``SW buildup'', $c_0=0.6\%$, $\theta_0=1.0^{\circ}$; and (c) ``SW growth'', $H=1.2\%$, $\theta_0=1.5^{\circ}$.} 
\end{figure}

We now move onto a study of the effect of initial lock concentration, $c_0$, on the development of the SW.  With respect to the previous section, we fix the current height at $H=1.5$ and study the dependance of the SW development on both $c_0$ and the slope angle, $\theta_0$.  It should be noted that changing $c_0$ has a direct effect on the system's time scale through $t_0=L_0/u_b$, where the buoyancy velocity is $u_b=\sqrt{gR_{*}c_0L_0}$.  In Fig. \ref{fig9} the substrate structure after 80 flows is shown for three different cases, each representative of one of the phases found in the previous section.  This figure is analogous to Fig. \ref{fig5} of the previous section.

\begin{figure}
\noindent\includegraphics[width=20pc]{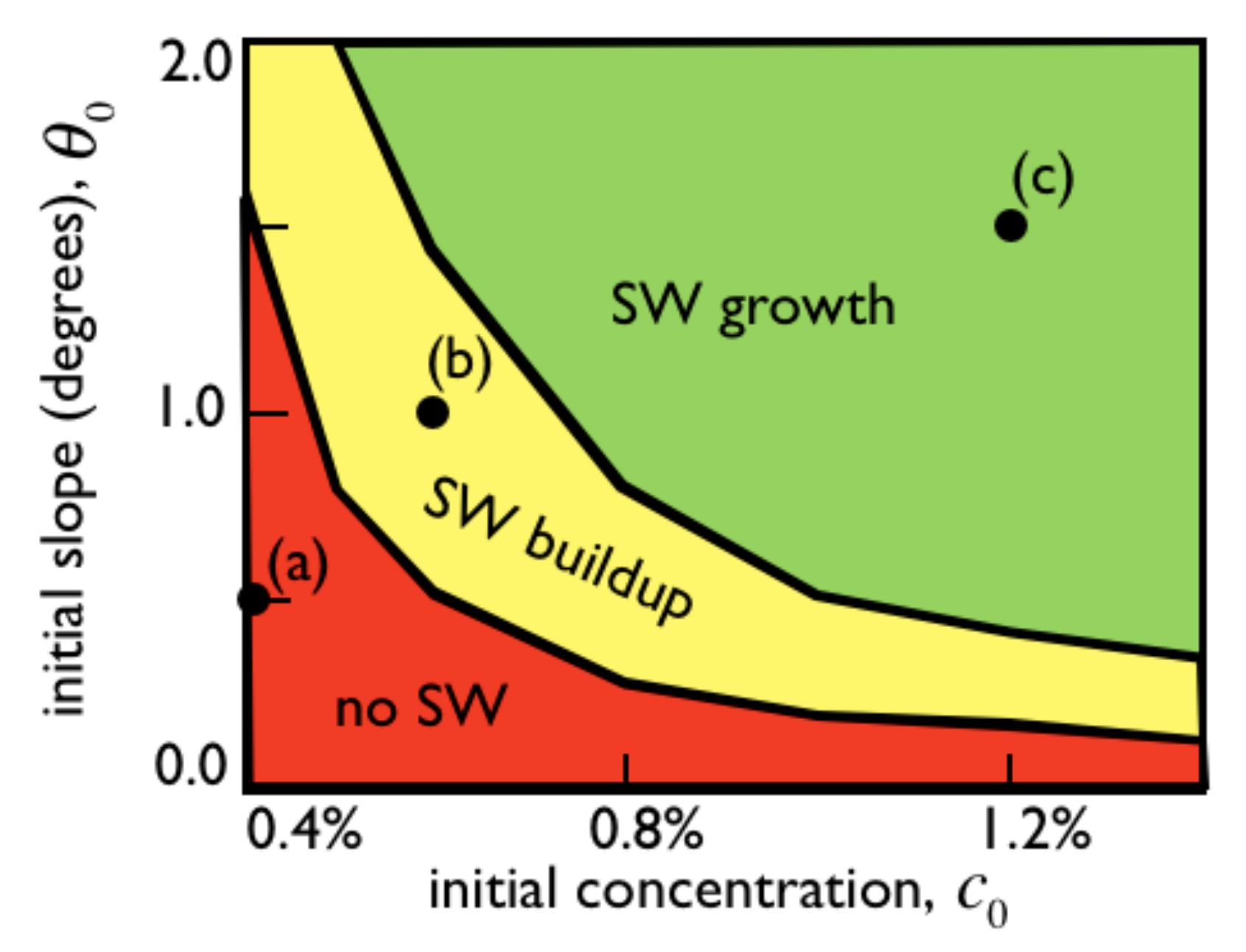}
\caption{\label{fig10} Phase diagram for SW in the $c_0$-$\theta_0$ plane, where $c_0$ is the initial particle concentration and $\theta_0$ is the initial slope of the substrate.  The three regions are identified according to the phase of the SW:  (red) ``no SW'', (yellow) ``SW buildup'', and (green) ``SW growth''.  The points simulated in Fig. \ref{fig9} are plotted as black dots and labeled (a, b, and c) consistent with the previous figure.} 
\end{figure}

A much larger set of simulations is used to define the the three regions corresponding to the phases of the SW, in a $c_0$-$\theta_0$ plane (where the initial lock height, $H$, and and particle size, $\{d_i\}$ are constants).  This phase diagram is shown in Fig. \ref{fig10}, and is analogous to Fig. \ref{fig8} of the previous section.  Two lines divide this plane into areas of ``no SW'', ``SW buildup'', and ``SW growth''.  As $c_0$ increases $\theta_0$ can be reduced and still maintain the SW phase.  The three exemplars shown in Fig. \ref{fig9} are indicated as black points on this figure.

\begin{figure}
\noindent\includegraphics[width=20pc]{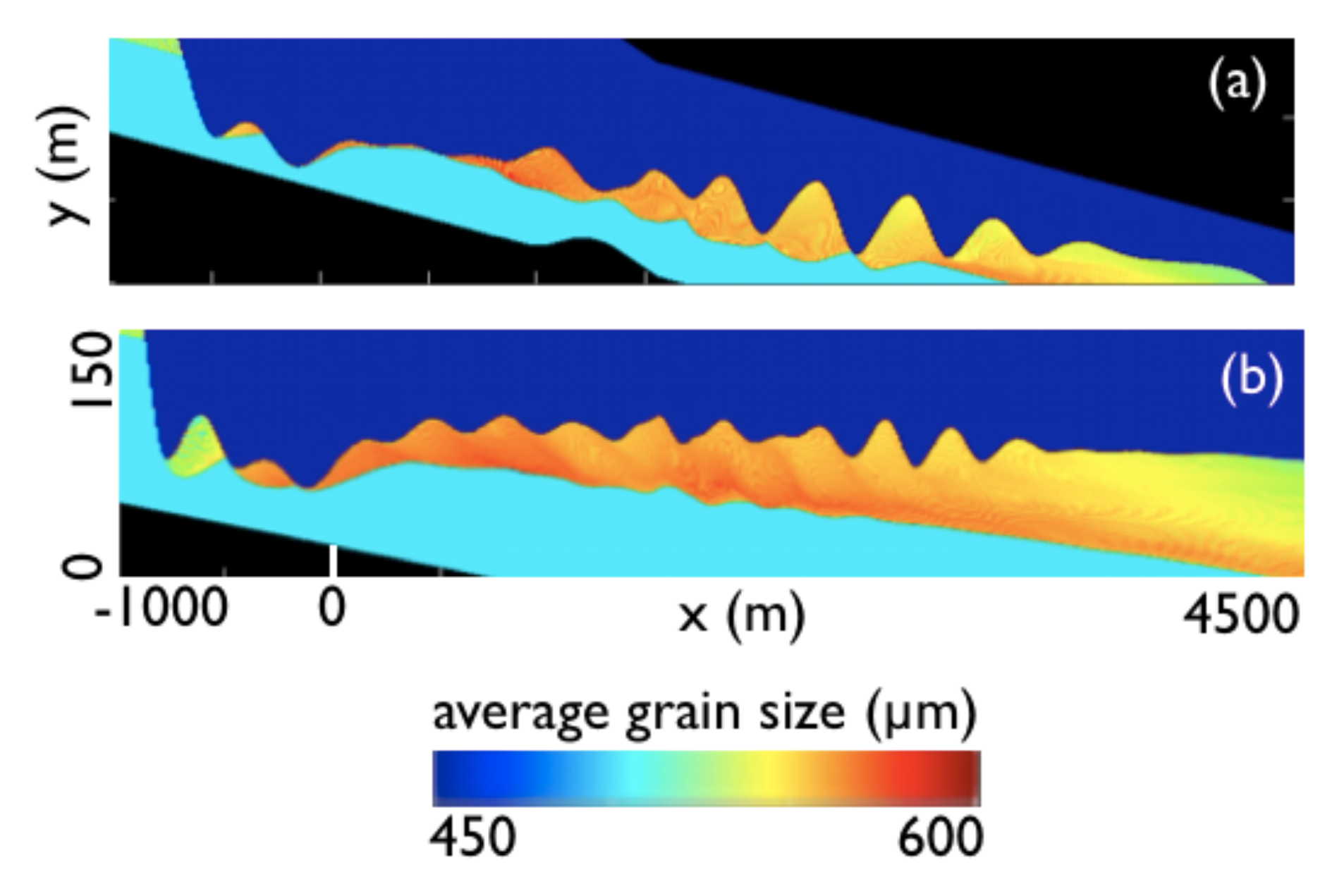}
\caption{\label{fig11} Development of SW on a substrate after 80 flows, to study the effect of the particle concentration, $c_0$, on the wavelength of the SW, $\lambda$.   The $x$-$y$ image is colored according to the average grain diameter.  The initial obstacle has the reduced height shown in Fig. \ref{fig4}.  The color bar is a rainbow, starting from 450 $\mu$m at blue and ending at 600 $\mu$m at red.  The profiles are shown for:  (a) $c_0=0.6\%$, $\theta_0=2.0^{\circ}$, where $\lambda=430$ m;  and (b) $c_0=1.2\%$, $\theta_0=1.5^{\circ}$, where $\lambda=310$ m.} 
\end{figure}

\begin{figure}
\noindent\includegraphics[width=20pc]{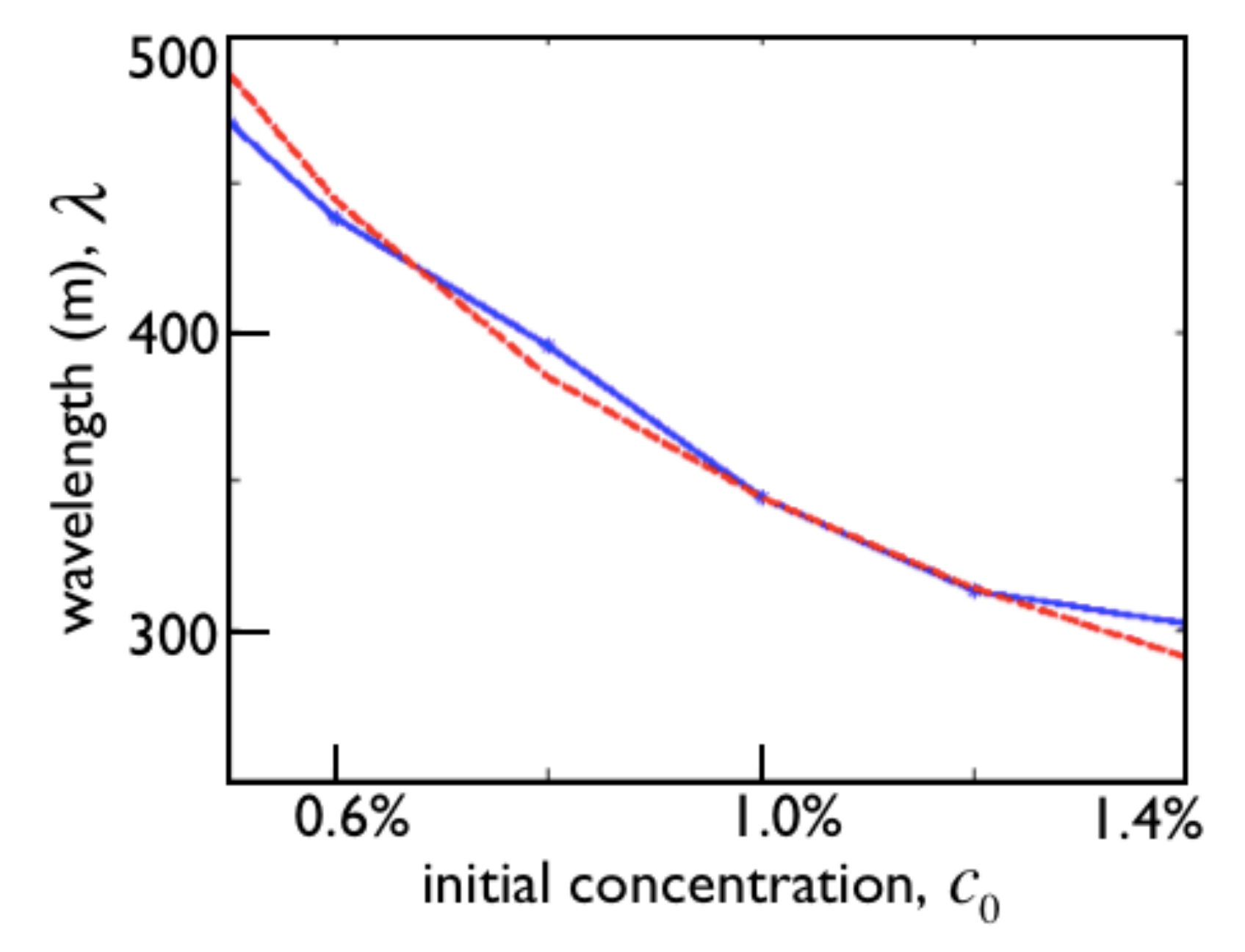}
\caption{\label{fig12} Dependance of SW wavelength, $\lambda$, on the initial particle concentration, $c_0$.  The result of a set of simulations similar to those shown in Fig. \ref{fig11} ($H=1.5$ and $\theta_0=1.5^{\circ}$) are shown as the blue line.  A fit to the data of the form $\lambda=\lambda_1 \sqrt{c_1/c_0}$, where $\lambda_1=344$ m and $c_1=1.0\%$, is shown as the dashed red line.}
\end{figure}

A closer look is taken at the dependance of the SW wavelength, $\lambda$, by studying two ``SW growth'' cases with different values of $c_0$.  These cases have $c_0$ values of 0.6\% and 1.2\%, and slope angles of 2$^{\circ}$ and 1.5$^{\circ}$, respectively.  The substrate structure after 80 flows is shown in Fig. \ref{fig11}.  Note that for the increase of $c_0$ by a factor of 2, the wavelength has decreased by a factor of $\sqrt{2}$ from 430 m to 310 m.  This is consistent with the decrease in the time scale by a factor $\sqrt{c_0}$ with the increase of $c_0$.  This dependance is further established by a larger set of simulations whose results are shown in Fig. \ref{fig12}.  Here the wavelength of the SW is plotted versus the initial concentration.  Notice the good fit of these points to a line of the form $\lambda \propto 1/\sqrt{c_0}$.  We also studied the effect on $\lambda$ of variation in the other controlling variables ($H$, $\theta_0$, and $\{d_i\}$).  We found that there was weak or little dependance on these variables.

\subsection{Dependance on particle diameter}
\begin{figure}
\noindent\includegraphics[width=20pc]{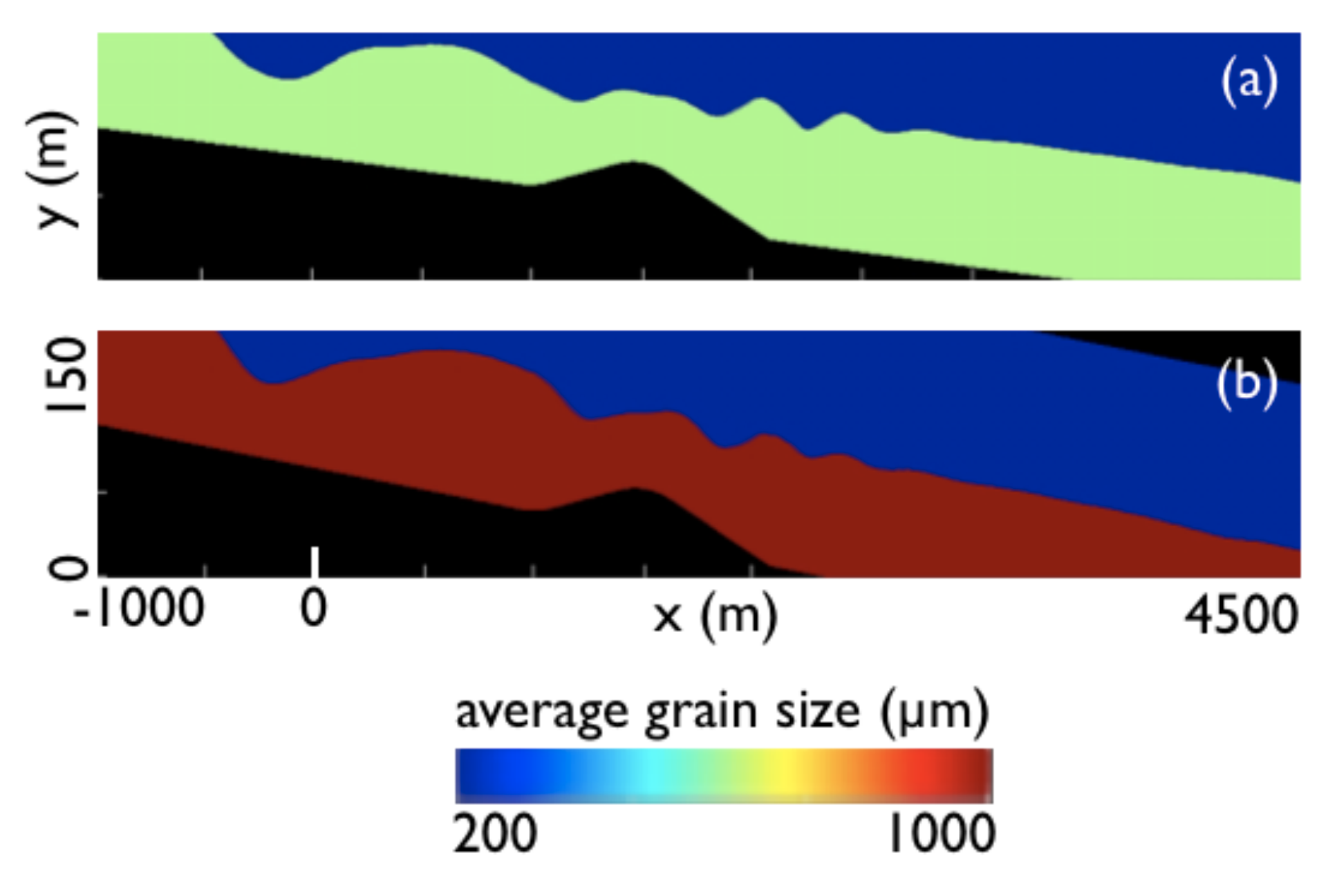}
\caption{\label{fig13} Development of SW on a substrate after 25 flows, to study the effect of the particle diameter, $d$.   The $x$-$y$ image is colored according to the average grain diameter.  The initial obstacle has the reduced height shown in Fig. \ref{fig4}.  The color bar is a rainbow, starting from 200 $\mu$m at blue and ending at 1000 $\mu$m at red.  The profiles are shown for:  (a) $d=600$ $\mu$m  ($R_{pi}=56$), $\theta_0=1.0^{\circ}$;  and (b) $d=1000$ $\mu$m ($R_{pi}=120$), $\theta_0=1.5^{\circ}$.} 
\end{figure}

\begin{figure}
\noindent\includegraphics[width=20pc]{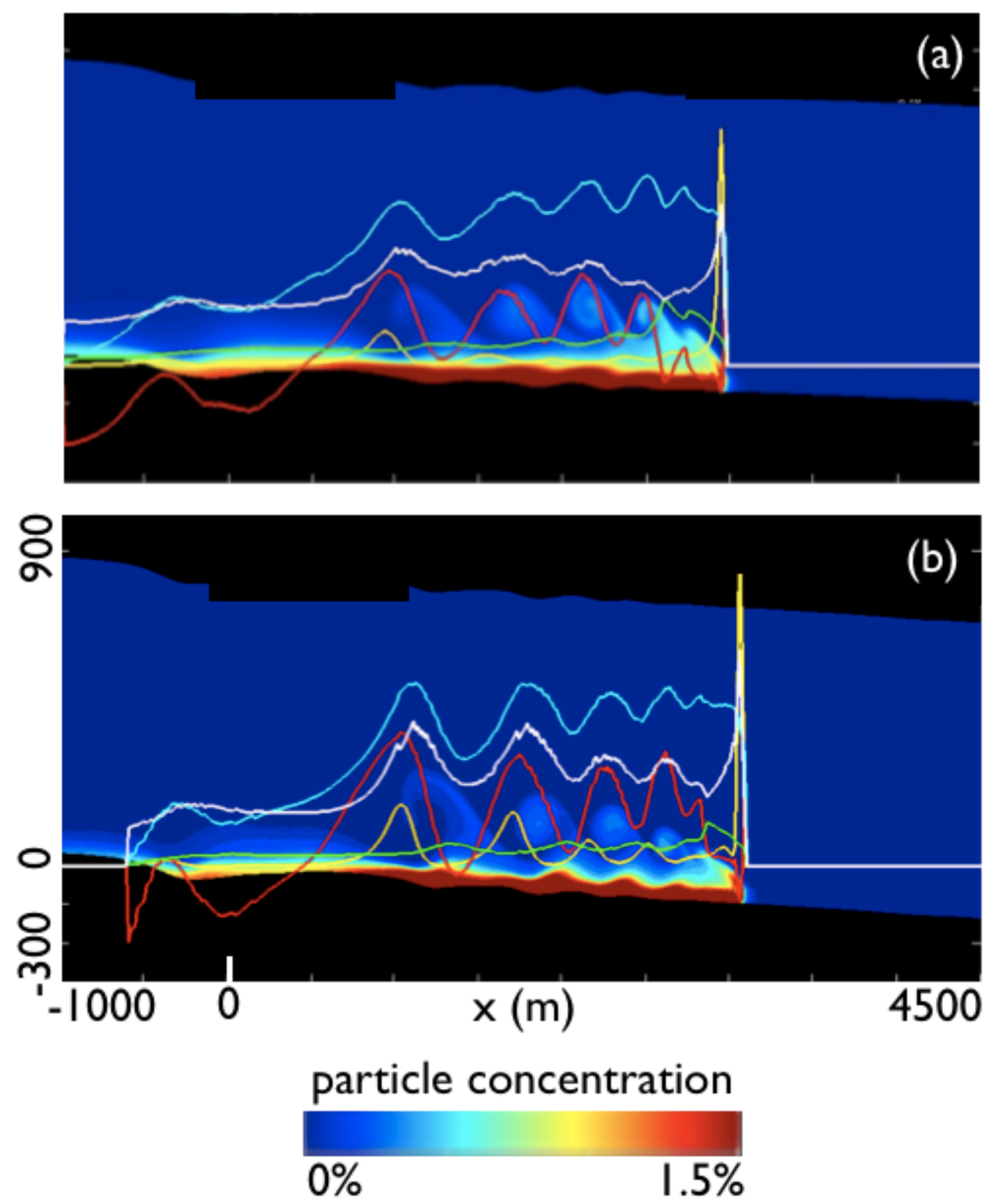}
\caption{\label{fig14} Particle volume concentration of the flow in the $x$-$y$ plane at the normalized time, $t=8$, for flow, $f=25$, to study the effect of the particle diameter, $d$.  This corresponds to simulations of Fig. \ref{fig13}.  The color bar is a rainbow, starting from 0 at blue and ending at 0.015 at red.  Also shown are the depth averaged current variables:  (blue) velocity $U \times 100$ in m/s, (white) concentration $C \times 2 \times 10^4$, (green) current height $h$ in m, (red) change in Froude number $F_2 \times 200$, and (yellow) shear velocity term in the resuspension $V_{\mathrm{shr}}^5 \times 5 \times 10^6$ in (m/s)$^5$.  All quantities plotted have SI dimensions.  The profiles are shown for:  (a) $d=600$ $\mu$m  ($R_{pi}=56$), $\theta_0=1.0^{\circ}$;  and (b) $d=1000$ $\mu$m  ($R_{pi}=120$), $\theta_0=1.5^{\circ}$.} 
\end{figure}

Finally, we turn our attention to the effect of particle diameter, $d$, on the development of the SW.  We fix the lock height at $H=1.5$, the number of grain types at one, and the initial particle concentration at $c_0=0.8\%$, and study the dependance of the SW development on both $d$ and the slope angle, $\theta_0$.  We present two cases in Figs. \ref{fig13} and \ref{fig14}, where we display the substrate structure after 25 flows.  The particle diameters are 600 $\mu$m and 1000 $\mu$m ($R_{pi}=56$ and 120), with slope angles of 1.0$^{\circ}$ and 1.5$^{\circ}$, respectively.  The two cases display a very similar development of ``SW growth'' to Figs. \ref{fig5}c and \ref{fig6}c.  As the grain diameter increases, the particle mass and the settling velocity  increases, leading to more difficult resuspension.  To obtain a similar SW growth for the larger grain diameter, the slope needs to be increased.

\begin{figure}
\noindent\includegraphics[width=20pc]{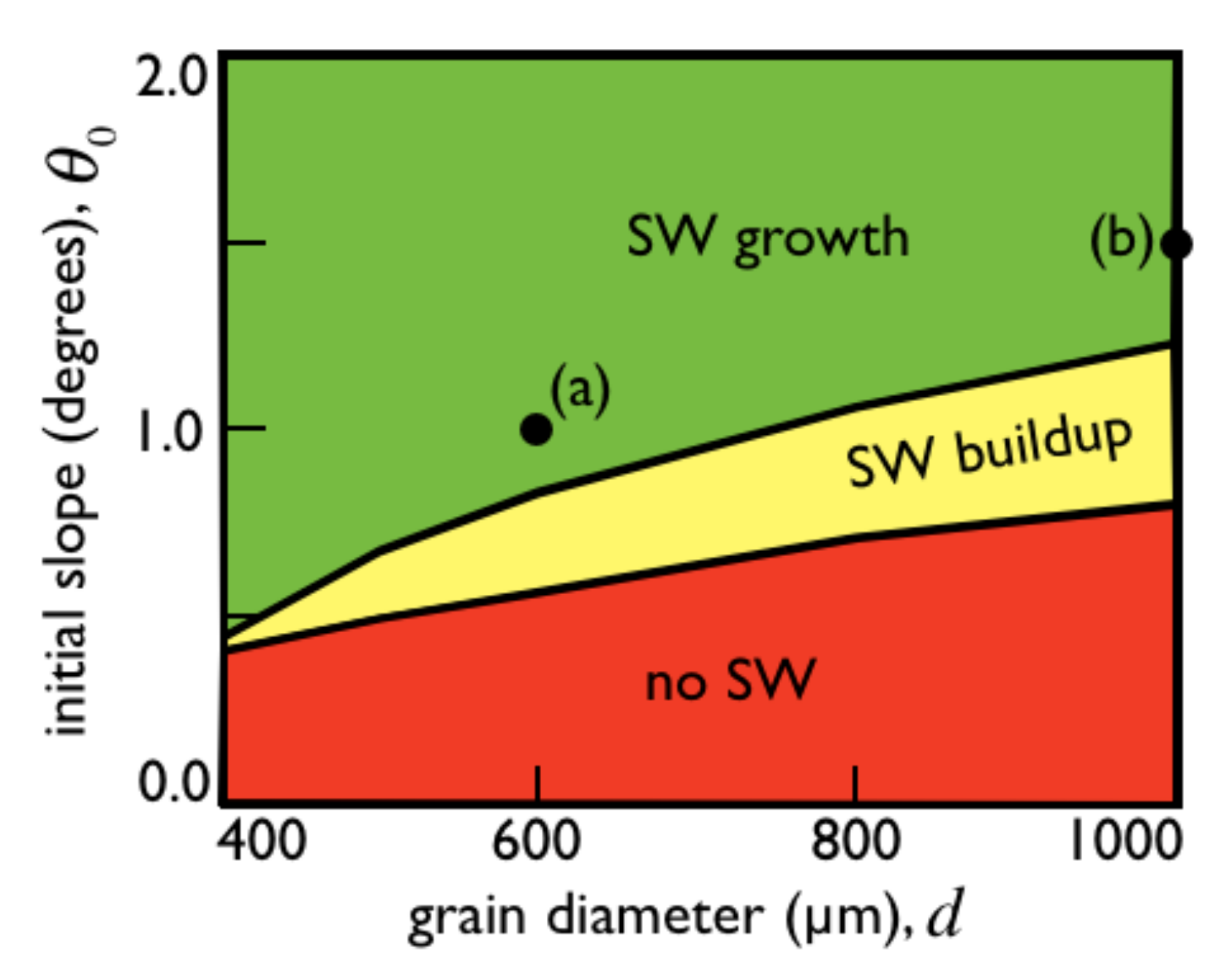}
\caption{\label{fig15} Phase diagram for SW in the $d$-$\theta_0$ plane, where $d$ is the particle diameter and $\theta_0$ is the initial slope of the substrate.  Range of $d$ displayed corresponds to $R_{pi}=30$ to 120.  The three regions are identified according to the phase of the SW:  (red) ``no SW'', (yellow) ``SW buildup'', and (green) ``SW growth''.  The points simulated in Fig. \ref{fig13} are plotted as green dots and labeled (a and b) consistent with the previous figure.} 
\end{figure}

A much larger set of simulations is used to define the the three regions corresponding to the phases of the SW, in a $d$-$\theta_0$ plane (where the initial lock height, $H$, and initial particle concentrations, $c_0$, are constants).  This phase diagram is shown in Fig. \ref{fig15}, and is analogous to Fig. \ref{fig10} of the previous section.  Two lines divide this plane into areas of ``no SW'', ``SW buildup'', and ``SW growth''.  The two exemplars shown in Fig. \ref{fig13} are indicated as black points on this figure.

\begin{figure}
\noindent\includegraphics[width=20pc]{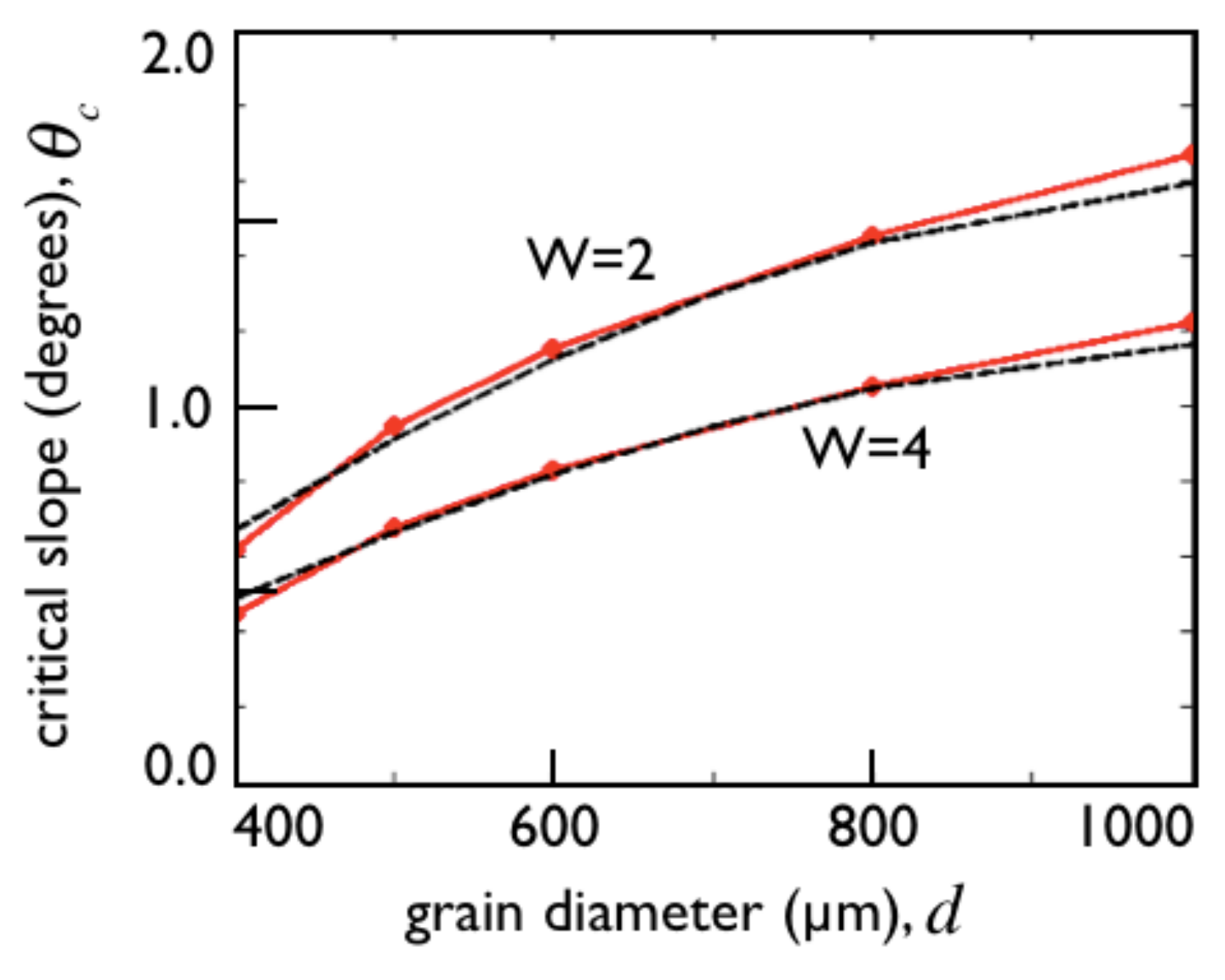}
\caption{\label{fig16} Critical angle for SW growth, $\theta_c$ as a function of grain diameter, $d$, for two different lock widths, $W$.  The $\theta_c(d)$ curves are shown a red solid lines.  A theoretical expression for the resuspension, $E_s$, is compared to these curves by plotting the function $A/E_s$, where $A=1.77$ fits the $W=4$ curve, and $A=2.42$ fits the $W=2$ curve.  These fit expressions are plotted as dotted black lines.} 
\end{figure}

To establish a further connection between SW generation and resuspension, systems with two different lock widths, $W$, were examined -- a width of 4 as in all previous simulations, and a reduced width of 2.  Fig. \ref{fig16} shows the change in the boundary of the ``SW growth'' phase in the $d$-$\theta_0$ plane with this decrease in $W$.  This boundary is given by the critical angle, $\theta_c$ as a function of grain diameter, $d$.  For this narrower lock, fewer particles are included in the current which increases the critical angle for the same diameter.  The two critical angle curves, $\theta_c(d)$, are compared to the normalized inverse of the resuspension term, $E_s$.  The Dietrich relation for the settling velocity with a characteristic normalized shear velocity of $V_{\mathrm{shr}}=0.15$ is used to calculate $E_s$.  The good correlation between these curves, $\theta_c(d)$ and $E_s^{-1}(d)$, shows that SW generation is highly correlated to the resuspension mechanism.

\section{Conclusions}
\label{conclusions}

After using a high resolution 2D computer simulation model of turbidity currents based on the Navier-Stokes equations, we have developed a more complete understanding of sediment wave generation.  This method took into account non-linearity, used a realistic erosion model, and modeled the depth dependant behaviour in a self consistent and self generating way.  The geometry was a lock release of a particle laden fluid onto a slope with a small obstacle.  Many flows were simulated, the next flow started after the previous flow had completed.  The obstacle is only a trigger for the sediment wave generation.  After several flows, the obstacle was eroded by the resuspension and a SW was generated, characterized by the most probably wavelength, $\lambda=2 \pi h$, derived by \citet{normark.etal.80}.  This is independent of any details of the initial obstacle.

The feedback mechanism responsible for the generation of SW comes from an interaction of the flow with the lower boundary condition.  This complex boundary condition modifies the topology of the boundary through the deposition of particles from the fluid and resuspension of particles from the substrate.  The increased slope on the downstream side of an obstacle increases the kinetic energy in a flow.  This will increase the resuspension, by increasing the shear in the fluid and the net effect will be increased erosion.  This erosion into the substrate, creates a subsequent decrease in slope.  As subsequent flows climb this decrease in slope, their kinetic energy decreases leading to increased deposition.  This creates another obstacle downdip of the original one.  The process then continues to generate a train of self generated and self consistent obstacles in the downstream direction.  This self consistent train of obstacles is the SW. 

There is an upward migration of the SW caused by another feedback mechanism.  Once the SW is established, the flow will preferentially deposit on the parts of the wave with increased slope and preferentially erode the parts of the wave with decreased slope.  The result will be a migration of the wave updip.
 
Conditions are not always favorable for having this feedback.  We found that there are four system parameters that influence the sediment wave growth:  (1) slope, $\theta_0$; (2) current lock height, $H$; (3) grain lock concentration, $c_0$; and (4) particle diameters, $\{d_i\}$.  Three phases of the system were found:  (1) ``no SW'', (2) ``SW buildup'', and (3) ``SW growth''.  These phases are characterized by whether or not the conditions are favorable for the feedback which leads to SW growth.  For the first phase, the conditions are always unfavorble.  For the second phase, the conditions are sometimes favorable (on the downslope side of the obstacle).  For the third phase, they are always favorable.  The conditions are determined by the parameters.  This allowed us to do systematic parameter studies and define three regions in the four dimensional $(\theta_0,H,c_0,d)$ space according to the phase of the system -- the phase diagram.

It should be noted why we only considered the four variables ($\theta_0$,$H$,$c_0$,$d$).  An analysis of dimensionless partial differential Eqs. (\ref{lie.derivative.vorticity})--(\ref{lie.derivative.c}) for $\omega$, $\psi$ and $\{c_i\}$ indicates that there should be three governing parameters associated with the gravity unit vector, $\hat{g}$, typical vorticity, $\omega_0$, and the average settling velocity, $\left<u_{si}\right>$.  Here we have reduced the set of particle concentration equations, over index $i$, to only one for the total concentration, $c$.  We have neglected first order, $\Delta d$ (sorting), and higher order effects on the substrate phase.   The dependance on $R_e$ and $P_e$ have been neglected because of the reasons stated in Sec. \ref{num.approach}.  Although the dependance on $c_0$ is normalized out of these equations, it is reintroduced by the resuspension in Eq. (\ref{garcia.parker.eqn}).  We now have four governing parameters.  The average $\left<u_{si}\right>$ can be associated with the particle size $d$, $\hat{g}$ can be associated with $\theta_0$, the resuspension with $c_0$, and finally $\omega_0$ can be associated with $H$.  The lock height, $H$, is really a surrogate for the flow size.  Little dependance was found to the aspect ratio, $W/H$, of the lock by \citet{blanchette.etal.05}, and the dependance on the lock width, $W$, can be normalized out of the problem by $L_o$.  We now see that the four governing parameters can be associated with the four variables that were studied.

Each phase was found to be characterized by several different things.  The first and most simple phase, ``no SW'', has a simple structure.  There in no development of SW or periodic structures in the flow.  The flow has a monotonically decreasing mass as a function of time.  There is no significant erosion.  The deposited substrate has little evidence of the individual flows.  It appears to have one massive bed that becomes gradually more fine grained downslope and gradually more coarse grained from bottom to top.  The second phase, ``SW buildup'', has some more structure.  There is a rather rapid local development of a SW, but this SW then reaches a steady state profile.  The flow, as a function of time, has a relatively constant mass with a maximum.  It shows a periodic structure in velocity, concentration and especially $F_2$ that correlates to the SW wave structure.  There is no appreciable erosion.  The deposited substrate still has little evidence of the individual flows.  It has one massive bed with gradually changing characteristics.  It becomes more fine grained downslope.  Vertically is shows more character than the first phase with the grain size showing an gradually oscillatory behavior.  The third phase, ``SW growth'', has significant structure.  There is a global development of a SW that initially grows exponentially.  The flow has a monotonically increasing mass that nearly doubles.  It shows structure in the velocity, concentration, $F_2$, and erosion that are sychronized to the SW structure.  It has significant erosion that increases exponentially in the upstream direction within the flow.  The deposited substrate has distinct deposited beds for each flow that show a complex structure.

We found that the driving force behind the establishment of the SW is the self sustainment of the flow.  This is evidenced by the time evolution of the mass, the threshold for the SW generation, boundaries of the SW phases, and the functional dependance of the critical angle, $\theta_0(d)$, for various initial lock widths.

The wavelength of the SW, $\lambda$, was found to be a significant function of the grain lock concentration, $c_0$.  It scaled as $1/\sqrt{c_0}$, directly related to how the time scales.  The three other system parameters were found to have weak or little effect on $\lambda$.

Finally, we discovered a rather direct path from the physics of the flow to the structure of the deposited substrate.  Starting with the flow, the work of \citet{blanchette.etal.05} established two regimes depending on whether the flow is self sustaining or depositing.  Our work has established the relationship between self sustainment within one flow and sediment wave generation on the substrate surface over multiple flows.  In fact, there are phases of SW formation, determined by whether the system never, sometimes or always generates SW.  The phase is determined by four system parameters, and a phase diagram can be constructed in terms of these parameters.  There are strong indications that there is a further direct relationship between the phase of SW formation and the structure of the deposited substrate.  This structure can be identified as a geologic texture, or more commonly called geologic facies.  This is a very remarkable result -- there are physical phases that could well correspond to geologic facies.  Future work will focus on further study on the emergent structure appearing in the deposited substrate.  Longer runs are needed to see the stationary character of the self organization, and modern techniques might be used to characterize the self organization.  We are also interested in understanding the effect of the sorting, $\Delta d$, on the substrate phase diagram.


%
%
%
%
%
%
%

\begin{acknowledgments}
The authors thank Eckart Meiburg, Ben Kneller and Francois Blanchette for many useful discussions.  We also are grateful to Stanislav Kuzmin for his help in developing the parallel computer code we used in the work and for his many useful technical interactions.  This work was supported by the CSIRO Office of the Chief Executive Science Fund.  Calculations were done using the CSIRO GPU Cluster with significant assistance from Ondrej Hlinka.
\end{acknowledgments}

%


\begin{thebibliography}{32}%
\makeatletter
\providecommand \@ifxundefined [1]{%
 \@ifx{#1\undefined}
}%
\providecommand \@ifnum [1]{%
 \ifnum #1\expandafter \@firstoftwo
 \else \expandafter \@secondoftwo
 \fi
}%
\providecommand \@ifx [1]{%
 \ifx #1\expandafter \@firstoftwo
 \else \expandafter \@secondoftwo
 \fi
}%
\providecommand \natexlab [1]{#1}%
\providecommand \enquote  [1]{``#1''}%
\providecommand \bibnamefont  [1]{#1}%
\providecommand \bibfnamefont [1]{#1}%
\providecommand \citenamefont [1]{#1}%
\providecommand \href@noop [0]{\@secondoftwo}%
\providecommand \href [0]{\begingroup \@sanitize@url \@href}%
\providecommand \@href[1]{\@@startlink{#1}\@@href}%
\providecommand \@@href[1]{\endgroup#1\@@endlink}%
\providecommand \@sanitize@url [0]{\catcode `\\12\catcode `\$12\catcode
  `\&12\catcode `\#12\catcode `\^12\catcode `\_12\catcode `\%12\relax}%
\providecommand \@@startlink[1]{}%
\providecommand \@@endlink[0]{}%
\providecommand \url  [0]{\begingroup\@sanitize@url \@url }%
\providecommand \@url [1]{\endgroup\@href {#1}{\urlprefix }}%
\providecommand \urlprefix  [0]{URL }%
\providecommand \Eprint [0]{\href }%
\@ifxundefined \urlstyle {%
  \providecommand \doi  [0]{\begingroup \@sanitize@url \@doi}%
  \providecommand \@doi [1]{\endgroup \@@startlink {\doibase
  #1}doi:\discretionary {}{}{}#1\@@endlink }%
}{%
  \providecommand \doi  [0]{doi:\discretionary{}{}{}\begingroup
  \urlstyle{rm}\Url }%
}%
\providecommand \doibase [0]{http://dx.doi.org/}%
\providecommand \Doi [0]{\begingroup \@sanitize@url \@Doi }%
\providecommand \@Doi  [1]{\endgroup\@@startlink{\doibase#1}\@@Doi}%
\providecommand \@@Doi [1]{#1\@@endlink}%
\providecommand \selectlanguage [0]{\@gobble}%
\providecommand \bibinfo  [0]{\@secondoftwo}%
\providecommand \bibfield  [0]{\@secondoftwo}%
\providecommand \translation [1]{[#1]}%
\providecommand \BibitemOpen [0]{}%
\providecommand \bibitemStop [0]{}%
\providecommand \bibitemNoStop [0]{.\EOS\space}%
\providecommand \EOS [0]{\spacefactor3000\relax}%
\providecommand \BibitemShut  [1]{\csname bibitem#1\endcsname}%
\bibitem [{\citenamefont {Meiburg}\ and\ \citenamefont
  {Kneller}(2010)}]{meiburg.kneller.10}%
  \BibitemOpen
  \bibfield  {author} {\bibinfo {author} {\bibfnamefont {E.}~\bibnamefont
  {Meiburg}}\ and\ \bibinfo {author} {\bibfnamefont {B.}~\bibnamefont
  {Kneller}},\ }\href@noop {} {\bibfield  {journal} {\bibinfo  {journal} {Ann.
  Rev. Fluid Mech.},\ }\textbf {\bibinfo {volume} {42}},\ \bibinfo {pages}
  {135} (\bibinfo {year} {2010})}\BibitemShut {NoStop}%
\bibitem [{\citenamefont {Parker}\ \emph {et~al.}(1986)\citenamefont {Parker},
  \citenamefont {Fukushima},\ and\ \citenamefont {Pantin}}]{parker.etal.86}%
  \BibitemOpen
  \bibfield  {author} {\bibinfo {author} {\bibfnamefont {G.}~\bibnamefont
  {Parker}}, \bibinfo {author} {\bibfnamefont {Y.}~\bibnamefont {Fukushima}}, \
  and\ \bibinfo {author} {\bibfnamefont {H.~M.}\ \bibnamefont {Pantin}},\
  }\href@noop {} {\bibfield  {journal} {\bibinfo  {journal} {J. Fluid Mech.},\
  }\textbf {\bibinfo {volume} {171}},\ \bibinfo {pages} {145} (\bibinfo {year}
  {1986})}\BibitemShut {NoStop}%
\bibitem [{\citenamefont {Blanchette}\ \emph {et~al.}(2005)\citenamefont
  {Blanchette}, \citenamefont {Strauss}, \citenamefont {Meiburg}, \citenamefont
  {Kneller},\ and\ \citenamefont {Glinsky}}]{blanchette.etal.05}%
  \BibitemOpen
  \bibfield  {author} {\bibinfo {author} {\bibfnamefont {F.}~\bibnamefont
  {Blanchette}}, \bibinfo {author} {\bibfnamefont {M.}~\bibnamefont {Strauss}},
  \bibinfo {author} {\bibfnamefont {E.}~\bibnamefont {Meiburg}}, \bibinfo
  {author} {\bibfnamefont {B.}~\bibnamefont {Kneller}}, \ and\ \bibinfo
  {author} {\bibfnamefont {M.~E.}\ \bibnamefont {Glinsky}},\ }\href@noop {}
  {\bibfield  {journal} {\bibinfo  {journal} {J. Geophys. Res.},\ }\textbf
  {\bibinfo {volume} {110}},\ \bibinfo {pages} {C120224} (\bibinfo {year}
  {2005})}\BibitemShut {NoStop}%
\bibitem [{\citenamefont {Sequeiros}\ \emph {et~al.}(2009)\citenamefont
  {Sequeiros}, \citenamefont {Naruse}, \citenamefont {Endo}, \citenamefont
  {Garcia},\ and\ \citenamefont {Parker}}]{sequeiros.etal.09}%
  \BibitemOpen
  \bibfield  {author} {\bibinfo {author} {\bibfnamefont {O.~E.}\ \bibnamefont
  {Sequeiros}}, \bibinfo {author} {\bibfnamefont {H.}~\bibnamefont {Naruse}},
  \bibinfo {author} {\bibfnamefont {N.}~\bibnamefont {Endo}}, \bibinfo {author}
  {\bibfnamefont {M.~H.}\ \bibnamefont {Garcia}}, \ and\ \bibinfo {author}
  {\bibfnamefont {G.}~\bibnamefont {Parker}},\ }\href@noop {} {\bibfield
  {journal} {\bibinfo  {journal} {J. Geophys. Res.},\ }\textbf {\bibinfo
  {volume} {114}},\ \bibinfo {pages} {C05025} (\bibinfo {year}
  {2009})}\BibitemShut {NoStop}%
\bibitem [{\citenamefont {Pantin}\ and\ \citenamefont
  {Franklin}(2009)}]{pantin.franklin.09}%
  \BibitemOpen
  \bibfield  {author} {\bibinfo {author} {\bibfnamefont {H.~M.}\ \bibnamefont
  {Pantin}}\ and\ \bibinfo {author} {\bibfnamefont {M.~C.}\ \bibnamefont
  {Franklin}},\ }\href@noop {} {\bibfield  {journal} {\bibinfo  {journal} {J.
  Sedim. Res.},\ }\textbf {\bibinfo {volume} {79}},\ \bibinfo {pages} {862}
  (\bibinfo {year} {2009})}\BibitemShut {NoStop}%
\bibitem [{\citenamefont {Wynn}\ \emph {et~al.}(2000)\citenamefont {Wynn},
  \citenamefont {Weaver}, \citenamefont {Ercilla}, \citenamefont {Stow},\ and\
  \citenamefont {Masson}}]{wynn.etal.00}%
  \BibitemOpen
  \bibfield  {author} {\bibinfo {author} {\bibfnamefont {R.~B.}\ \bibnamefont
  {Wynn}}, \bibinfo {author} {\bibfnamefont {P.~P.~E.}\ \bibnamefont {Weaver}},
  \bibinfo {author} {\bibfnamefont {G.}~\bibnamefont {Ercilla}}, \bibinfo
  {author} {\bibfnamefont {D.~A.~V.}\ \bibnamefont {Stow}}, \ and\ \bibinfo
  {author} {\bibfnamefont {D.~G.}\ \bibnamefont {Masson}},\ }\href@noop {}
  {\bibfield  {journal} {\bibinfo  {journal} {Sedimentology},\ }\textbf
  {\bibinfo {volume} {47}},\ \bibinfo {pages} {1181} (\bibinfo {year}
  {2000})}\BibitemShut {NoStop}%
\bibitem [{\citenamefont {Wynn}\ and\ \citenamefont
  {Stow}(2002)}]{wynn.stow.02}%
  \BibitemOpen
  \bibfield  {author} {\bibinfo {author} {\bibfnamefont {R.~B.}\ \bibnamefont
  {Wynn}}\ and\ \bibinfo {author} {\bibfnamefont {D.~A.~V.}\ \bibnamefont
  {Stow}},\ }\href@noop {} {\bibfield  {journal} {\bibinfo  {journal} {Marine
  and Geology},\ }\textbf {\bibinfo {volume} {192}},\ \bibinfo {pages} {7}
  (\bibinfo {year} {2002})}\BibitemShut {NoStop}%
\bibitem [{\citenamefont {Kubo}\ and\ \citenamefont
  {Nakajima}(2002)}]{kubo.nakajima.02}%
  \BibitemOpen
  \bibfield  {author} {\bibinfo {author} {\bibfnamefont {Y.}~\bibnamefont
  {Kubo}}\ and\ \bibinfo {author} {\bibfnamefont {T.}~\bibnamefont
  {Nakajima}},\ }\href@noop {} {\bibfield  {journal} {\bibinfo  {journal}
  {Marine Geology},\ }\textbf {\bibinfo {volume} {192}},\ \bibinfo {pages}
  {105} (\bibinfo {year} {2002})}\BibitemShut {NoStop}%
\bibitem [{\citenamefont {Lee}\ \emph {et~al.}(2002)\citenamefont {Lee},
  \citenamefont {Syvitski}, \citenamefont {Parker}, \citenamefont {Orange},
  \citenamefont {Locat}, \citenamefont {Hutton},\ and\ \citenamefont
  {Imran}}]{lee.etal.02}%
  \BibitemOpen
  \bibfield  {author} {\bibinfo {author} {\bibfnamefont {H.~J.}\ \bibnamefont
  {Lee}}, \bibinfo {author} {\bibfnamefont {J.~P.~M.}\ \bibnamefont
  {Syvitski}}, \bibinfo {author} {\bibfnamefont {G.}~\bibnamefont {Parker}},
  \bibinfo {author} {\bibfnamefont {D.}~\bibnamefont {Orange}}, \bibinfo
  {author} {\bibfnamefont {J.}~\bibnamefont {Locat}}, \bibinfo {author}
  {\bibfnamefont {E.~W.~H.}\ \bibnamefont {Hutton}}, \ and\ \bibinfo {author}
  {\bibfnamefont {J.}~\bibnamefont {Imran}},\ }\href@noop {} {\bibfield
  {journal} {\bibinfo  {journal} {Marine Geology},\ }\textbf {\bibinfo {volume}
  {192}},\ \bibinfo {pages} {79} (\bibinfo {year} {2002})}\BibitemShut
  {NoStop}%
\bibitem [{\citenamefont {Normark}\ \emph {et~al.}(1980)\citenamefont
  {Normark}, \citenamefont {Hess}, \citenamefont {Stow},\ and\ \citenamefont
  {Bowen}}]{normark.etal.80}%
  \BibitemOpen
  \bibfield  {author} {\bibinfo {author} {\bibfnamefont {W.~R.}\ \bibnamefont
  {Normark}}, \bibinfo {author} {\bibfnamefont {G.~R.}\ \bibnamefont {Hess}},
  \bibinfo {author} {\bibfnamefont {D.~A.~V.}\ \bibnamefont {Stow}}, \ and\
  \bibinfo {author} {\bibfnamefont {A.~J.}\ \bibnamefont {Bowen}},\ }\href@noop
  {} {\bibfield  {journal} {\bibinfo  {journal} {Marine Geology},\ }\textbf
  {\bibinfo {volume} {37}},\ \bibinfo {pages} {1} (\bibinfo {year}
  {1980})}\BibitemShut {NoStop}%
\bibitem [{\citenamefont {Parker}\ and\ \citenamefont
  {Izumi}(2000)}]{parker.izumi.00}%
  \BibitemOpen
  \bibfield  {author} {\bibinfo {author} {\bibfnamefont {G.}~\bibnamefont
  {Parker}}\ and\ \bibinfo {author} {\bibfnamefont {N.}~\bibnamefont {Izumi}},\
  }\href@noop {} {\bibfield  {journal} {\bibinfo  {journal} {J. Fluid Mech.},\
  }\textbf {\bibinfo {volume} {419}},\ \bibinfo {pages} {203} (\bibinfo {year}
  {2000})}\BibitemShut {NoStop}%
\bibitem [{\citenamefont {Taki}\ and\ \citenamefont
  {Parker}(2005)}]{taki.parker.05}%
  \BibitemOpen
  \bibfield  {author} {\bibinfo {author} {\bibfnamefont {K.}~\bibnamefont
  {Taki}}\ and\ \bibinfo {author} {\bibfnamefont {G.}~\bibnamefont {Parker}},\
  }\href@noop {} {\bibfield  {journal} {\bibinfo  {journal} {J. Hydraul.
  Res.},\ }\textbf {\bibinfo {volume} {43}},\ \bibinfo {pages} {488} (\bibinfo
  {year} {2005})}\BibitemShut {NoStop}%
\bibitem [{\citenamefont {Sun}\ and\ \citenamefont
  {Parker}(2005)}]{sun.parker.05}%
  \BibitemOpen
  \bibfield  {author} {\bibinfo {author} {\bibfnamefont {T.}~\bibnamefont
  {Sun}}\ and\ \bibinfo {author} {\bibfnamefont {G.}~\bibnamefont {Parker}},\
  }\href@noop {} {\bibfield  {journal} {\bibinfo  {journal} {J. Hydraul.
  Res.},\ }\textbf {\bibinfo {volume} {43}},\ \bibinfo {pages} {502} (\bibinfo
  {year} {2005})}\BibitemShut {NoStop}%
\bibitem [{\citenamefont {Fildani}\ \emph {et~al.}(2006)\citenamefont
  {Fildani}, \citenamefont {Normark}, \citenamefont {Kostic},\ and\
  \citenamefont {Parker}}]{fildani.etal.06}%
  \BibitemOpen
  \bibfield  {author} {\bibinfo {author} {\bibfnamefont {A.}~\bibnamefont
  {Fildani}}, \bibinfo {author} {\bibfnamefont {W.~R.}\ \bibnamefont
  {Normark}}, \bibinfo {author} {\bibfnamefont {S.}~\bibnamefont {Kostic}}, \
  and\ \bibinfo {author} {\bibfnamefont {G.}~\bibnamefont {Parker}},\
  }\href@noop {} {\bibfield  {journal} {\bibinfo  {journal} {Sedimentology},\
  }\textbf {\bibinfo {volume} {53}},\ \bibinfo {pages} {1265} (\bibinfo {year}
  {2006})}\BibitemShut {NoStop}%
\bibitem [{\citenamefont {Kostic}\ and\ \citenamefont
  {Parker}(2006)}]{kostic.parker.06}%
  \BibitemOpen
  \bibfield  {author} {\bibinfo {author} {\bibfnamefont {S.}~\bibnamefont
  {Kostic}}\ and\ \bibinfo {author} {\bibfnamefont {G.}~\bibnamefont
  {Parker}},\ }\href@noop {} {\bibfield  {journal} {\bibinfo  {journal} {J.
  Hydraul. Res.},\ }\textbf {\bibinfo {volume} {44}},\ \bibinfo {pages} {631}
  (\bibinfo {year} {2006})}\BibitemShut {NoStop}%
\bibitem [{\citenamefont {Hall}\ \emph {et~al.}(2008)\citenamefont {Hall},
  \citenamefont {Meiburg},\ and\ \citenamefont {Kneller}}]{hall.etal.08}%
  \BibitemOpen
  \bibfield  {author} {\bibinfo {author} {\bibfnamefont {B.}~\bibnamefont
  {Hall}}, \bibinfo {author} {\bibfnamefont {E.}~\bibnamefont {Meiburg}}, \
  and\ \bibinfo {author} {\bibfnamefont {B.}~\bibnamefont {Kneller}},\
  }\href@noop {} {\bibfield  {journal} {\bibinfo  {journal} {J. Fluid Mech.},\
  }\textbf {\bibinfo {volume} {615}},\ \bibinfo {pages} {185} (\bibinfo {year}
  {2008})}\BibitemShut {NoStop}%
\bibitem [{\citenamefont {Hall}(2009)}]{hall.09}%
  \BibitemOpen
  \bibfield  {author} {\bibinfo {author} {\bibfnamefont {B.}~\bibnamefont
  {Hall}},\ }\emph {\bibinfo {title} {Submarine channel and sediment wave
  formation by turbidity currents: Navier-Stokes based linear stability
  analysis}},\ \href@noop {} {\bibinfo {type} {{Ph.D.} thesis}},\ \bibinfo
  {school} {UCSB}, \bibinfo {address} {Santa Barbara, CA} (\bibinfo {year}
  {2009})\BibitemShut {NoStop}%
\bibitem [{\citenamefont {Hartel}\ \emph {et~al.}(2000)\citenamefont {Hartel},
  \citenamefont {Meiburg},\ and\ \citenamefont {Necker}}]{hartel.etal.00}%
  \BibitemOpen
  \bibfield  {author} {\bibinfo {author} {\bibfnamefont {C.}~\bibnamefont
  {Hartel}}, \bibinfo {author} {\bibfnamefont {E.}~\bibnamefont {Meiburg}}, \
  and\ \bibinfo {author} {\bibfnamefont {F.}~\bibnamefont {Necker}},\
  }\href@noop {} {\bibfield  {journal} {\bibinfo  {journal} {J. Fluid Mech.},\
  }\textbf {\bibinfo {volume} {418}},\ \bibinfo {pages} {189} (\bibinfo {year}
  {2000})}\BibitemShut {NoStop}%
\bibitem [{\citenamefont {Necker}\ \emph {et~al.}(2001)\citenamefont {Necker},
  \citenamefont {Hartel}, \citenamefont {Kleiser},\ and\ \citenamefont
  {Meiburg}}]{necker.etal.02}%
  \BibitemOpen
  \bibfield  {author} {\bibinfo {author} {\bibfnamefont {F.}~\bibnamefont
  {Necker}}, \bibinfo {author} {\bibfnamefont {C.}~\bibnamefont {Hartel}},
  \bibinfo {author} {\bibfnamefont {L.}~\bibnamefont {Kleiser}}, \ and\
  \bibinfo {author} {\bibfnamefont {E.}~\bibnamefont {Meiburg}},\ }\href@noop
  {} {\bibfield  {journal} {\bibinfo  {journal} {Int. J. Multiphase Flow},\
  }\textbf {\bibinfo {volume} {28}},\ \bibinfo {pages} {279} (\bibinfo {year}
  {2001})}\BibitemShut {NoStop}%
\bibitem [{\citenamefont {Garcia}\ and\ \citenamefont
  {Parker}(1993)}]{garcia.parker.93}%
  \BibitemOpen
  \bibfield  {author} {\bibinfo {author} {\bibfnamefont {M.~H.}\ \bibnamefont
  {Garcia}}\ and\ \bibinfo {author} {\bibfnamefont {G.}~\bibnamefont
  {Parker}},\ }\href@noop {} {\bibfield  {journal} {\bibinfo  {journal} {J.
  Geophys. Res.},\ }\textbf {\bibinfo {volume} {98}},\ \bibinfo {pages} {4793}
  (\bibinfo {year} {1993})}\BibitemShut {NoStop}%
\bibitem [{\citenamefont {Wright}\ and\ \citenamefont
  {Parker}(2004)}]{wright.parker.04}%
  \BibitemOpen
  \bibfield  {author} {\bibinfo {author} {\bibfnamefont {S.}~\bibnamefont
  {Wright}}\ and\ \bibinfo {author} {\bibfnamefont {G.}~\bibnamefont
  {Parker}},\ }\href@noop {} {\bibfield  {journal} {\bibinfo  {journal} {J.
  Hydraul. Eng.},\ }\textbf {\bibinfo {volume} {130}},\ \bibinfo {pages} {796}
  (\bibinfo {year} {2004})}\BibitemShut {NoStop}%
\bibitem [{\citenamefont {Ablowitz}\ and\ \citenamefont
  {Segur}(1984)}]{ablowitz.segur.84}%
  \BibitemOpen
  \bibfield  {author} {\bibinfo {author} {\bibfnamefont {M.~J.}\ \bibnamefont
  {Ablowitz}}\ and\ \bibinfo {author} {\bibfnamefont {H.}~\bibnamefont
  {Segur}},\ }\href@noop {} {\emph {\bibinfo {title} {Solitons and the Inverse
  Scattering Transform}}}\ (\bibinfo  {publisher} {Society for Industrial and
  Applied Mathematics},\ \bibinfo {address} {Philadelphia, PA},\ \bibinfo
  {year} {1984})\BibitemShut {NoStop}%
\bibitem [{\citenamefont {Arecchi}\ \emph {et~al.}(1989)\citenamefont
  {Arecchi}, \citenamefont {Gadomski}, \citenamefont {Meucci},\ and\
  \citenamefont {Roversi}}]{arecchi.etal.89}%
  \BibitemOpen
  \bibfield  {author} {\bibinfo {author} {\bibfnamefont {F.~T.}\ \bibnamefont
  {Arecchi}}, \bibinfo {author} {\bibfnamefont {W.}~\bibnamefont {Gadomski}},
  \bibinfo {author} {\bibfnamefont {R.}~\bibnamefont {Meucci}}, \ and\ \bibinfo
  {author} {\bibfnamefont {J.~A.}\ \bibnamefont {Roversi}},\ }\href@noop {}
  {\bibfield  {journal} {\bibinfo  {journal} {Phys. Rev. A},\ }\textbf
  {\bibinfo {volume} {39}},\ \bibinfo {pages} {4004} (\bibinfo {year}
  {1989})}\BibitemShut {NoStop}%
\bibitem [{\citenamefont {Blanchette}\ \emph {et~al.}(2006)\citenamefont
  {Blanchette}, \citenamefont {Piche}, \citenamefont {Meiburg},\ and\
  \citenamefont {Strauss}}]{blanchette.etal.06}%
  \BibitemOpen
  \bibfield  {author} {\bibinfo {author} {\bibfnamefont {F.}~\bibnamefont
  {Blanchette}}, \bibinfo {author} {\bibfnamefont {V.}~\bibnamefont {Piche}},
  \bibinfo {author} {\bibfnamefont {E.}~\bibnamefont {Meiburg}}, \ and\
  \bibinfo {author} {\bibfnamefont {M.}~\bibnamefont {Strauss}},\ }\href@noop
  {} {\bibfield  {journal} {\bibinfo  {journal} {Computers and Fluids},\
  }\textbf {\bibinfo {volume} {35}},\ \bibinfo {pages} {492} (\bibinfo {year}
  {2006})}\BibitemShut {NoStop}%
\bibitem [{\citenamefont {Parker}\ \emph {et~al.}(2000)\citenamefont {Parker},
  \citenamefont {Parola},\ and\ \citenamefont {Leclair}}]{parker.etal.00}%
  \BibitemOpen
  \bibfield  {author} {\bibinfo {author} {\bibfnamefont {G.}~\bibnamefont
  {Parker}}, \bibinfo {author} {\bibfnamefont {C.}~\bibnamefont {Parola}}, \
  and\ \bibinfo {author} {\bibfnamefont {S.}~\bibnamefont {Leclair}},\
  }\href@noop {} {\bibfield  {journal} {\bibinfo  {journal} {J. Hydraul.
  Eng.},\ }\textbf {\bibinfo {volume} {126}},\ \bibinfo {pages} {11} (\bibinfo
  {year} {2000})}\BibitemShut {NoStop}%
\bibitem [{\citenamefont {Pratson}\ \emph {et~al.}(2001)\citenamefont
  {Pratson}, \citenamefont {Imran}, \citenamefont {Hutton}, \citenamefont
  {Parker},\ and\ \citenamefont {Syvitski}}]{pratson.etal.01}%
  \BibitemOpen
  \bibfield  {author} {\bibinfo {author} {\bibfnamefont {L.~F.}\ \bibnamefont
  {Pratson}}, \bibinfo {author} {\bibfnamefont {J.}~\bibnamefont {Imran}},
  \bibinfo {author} {\bibfnamefont {E.~W.~H.}\ \bibnamefont {Hutton}}, \bibinfo
  {author} {\bibfnamefont {G.}~\bibnamefont {Parker}}, \ and\ \bibinfo {author}
  {\bibfnamefont {J.~P.~M.}\ \bibnamefont {Syvitski}},\ }\href@noop {}
  {\bibfield  {journal} {\bibinfo  {journal} {Computers and Geosciences},\
  }\textbf {\bibinfo {volume} {27}},\ \bibinfo {pages} {701} (\bibinfo {year}
  {2001})}\BibitemShut {NoStop}%
\bibitem [{\citenamefont {Dietrich}(1982)}]{dietrich.82}%
  \BibitemOpen
  \bibfield  {author} {\bibinfo {author} {\bibfnamefont {W.~E.}\ \bibnamefont
  {Dietrich}},\ }\href@noop {} {\bibfield  {journal} {\bibinfo  {journal}
  {Water Resources},\ }\textbf {\bibinfo {volume} {18}},\ \bibinfo {pages}
  {1615} (\bibinfo {year} {1982})}\BibitemShut {NoStop}%
\bibitem [{\citenamefont {Felix}(2001)}]{felix.01}%
  \BibitemOpen
  \bibfield  {author} {\bibinfo {author} {\bibfnamefont {M.}~\bibnamefont
  {Felix}},\ }\href@noop {} {\bibfield  {journal} {\bibinfo  {journal} {Spec.
  Publs. Int. Ass. Sediment},\ }\textbf {\bibinfo {volume} {31}},\ \bibinfo
  {pages} {71} (\bibinfo {year} {2001})}\BibitemShut {NoStop}%
\bibitem [{\citenamefont {Choi}\ and\ \citenamefont
  {Garcia}(2002)}]{choi.garcia.02}%
  \BibitemOpen
  \bibfield  {author} {\bibinfo {author} {\bibfnamefont {S.}~\bibnamefont
  {Choi}}\ and\ \bibinfo {author} {\bibfnamefont {M.~H.}\ \bibnamefont
  {Garcia}},\ }\href@noop {} {\bibfield  {journal} {\bibinfo  {journal} {J.
  Hydral. Eng.},\ }\textbf {\bibinfo {volume} {128}},\ \bibinfo {pages} {55}
  (\bibinfo {year} {2002})}\BibitemShut {NoStop}%
\bibitem [{\citenamefont {Huang}\ \emph {et~al.}(2000)\citenamefont {Huang},
  \citenamefont {Imran},\ and\ \citenamefont {Pirmez}}]{huang.etal.08}%
  \BibitemOpen
  \bibfield  {author} {\bibinfo {author} {\bibfnamefont {H.}~\bibnamefont
  {Huang}}, \bibinfo {author} {\bibfnamefont {J.}~\bibnamefont {Imran}}, \ and\
  \bibinfo {author} {\bibfnamefont {C.}~\bibnamefont {Pirmez}},\ }\href@noop {}
  {\bibfield  {journal} {\bibinfo  {journal} {J. Hydraul. Eng.},\ }\textbf
  {\bibinfo {volume} {134}},\ \bibinfo {pages} {621} (\bibinfo {year}
  {2000})}\BibitemShut {NoStop}%
\bibitem [{\citenamefont {Lele}(1992)}]{lele.92}%
  \BibitemOpen
  \bibfield  {author} {\bibinfo {author} {\bibfnamefont {S.~K.}\ \bibnamefont
  {Lele}},\ }\href@noop {} {\bibfield  {journal} {\bibinfo  {journal} {J. Comp.
  Phys.},\ }\textbf {\bibinfo {volume} {103}},\ \bibinfo {pages} {16} (\bibinfo
  {year} {1992})}\BibitemShut {NoStop}%
\bibitem [{\citenamefont {Middleton}(1993)}]{middleton.93}%
  \BibitemOpen
  \bibfield  {author} {\bibinfo {author} {\bibfnamefont {G.~V.}\ \bibnamefont
  {Middleton}},\ }\href@noop {} {\bibfield  {journal} {\bibinfo  {journal}
  {Earth Planet Sci.},\ }\textbf {\bibinfo {volume} {21}},\ \bibinfo {pages}
  {89} (\bibinfo {year} {1993})}\BibitemShut {NoStop}%
\end{thebibliography}

%

\end{document}